\documentclass[twocolumn]{aastex631}

\usepackage{hyperref}
\usepackage{verbatim}
\usepackage{graphicx,url}
\usepackage{amsmath,amssymb}
\usepackage{hyperref}
\hypersetup{colorlinks,
	linkcolor=blue, 
	urlcolor=magenta, 
	anchorcolor=blue,
	citecolor=blue
}
\usepackage{subfigure}

\begin{document}

\title{CMBFSCNN: Cosmic Microwave Background Polarization Foreground Subtraction with Convolutional Neural Network}


\author{Ye-Peng Yan}
\affil{Institute for Frontiers in Astronomy and Astrophysics, Beijing Normal University, Beijing 100875, China; xiajq@bnu.edu.cn}
\affil{Key Laboratory of Particle Astrophysics, Institute of High Energy Physics, Chinese Academy of Science, P. O. Box 918-3 Beijing 100049, People’s Republic of China}
\affil{Department of Astronomy, Beijing Normal University, Beijing 100875, China}

\author{Si-Yu Li}
\affil{Key Laboratory of Particle Astrophysics, Institute of High Energy Physics, Chinese Academy of Science, P. O. Box 918-3 Beijing 100049, People’s Republic of China}

\author{Guo-Jian Wang}
\affil{Department of Physics, Stellenbosch University, Matieland 7602, South Africa}
\affil{National Institute for Theoretical and Computational Sciences (NITheCS)}

\author{Zirui Zhang}
\affil{Institute of Frontier and Interdisciplinary Science and Key Laboratory of Particle Physics and Particle Irradiation (MOE), Shandong University, Qingdao 266237, China}
\affil{Key Laboratory of Particle Astrophysics, Institute of High Energy Physics, Chinese Academy of Science, P. O. Box 918-3 Beijing 100049, People’s Republic of China}

\author{Jun-Qing Xia}
\affil{Institute for Frontiers in Astronomy and Astrophysics, Beijing Normal University, Beijing 100875, China; xiajq@bnu.edu.cn}
\affil{Department of Astronomy, Beijing Normal University, Beijing 100875, China}

\begin{abstract}
In our previous study, we introduced a machine-learning technique, namely \texttt{CMBFSCNN}, for the removal of foreground contamination in cosmic microwave background (CMB) polarization data. This method was successfully employed on actual observational data from the Planck mission. In this study, we extend our investigation by considering the CMB lensing effect in simulated data and utilizing the \texttt{CMBFSCNN} approach to recover the CMB lensing B-mode power spectrum from multi-frequency observational maps. Our method is first applied to simulated data with the performance of CMB-S4 experiment. We achieve reliable recovery of the noisy CMB Q (or U) maps with a mean absolute difference of $0.016\pm0.008\ \mu$K (or $0.021\pm0.002\ \mu$K) for the CMB-S4 experiment. To address the residual instrumental noise in the foreground-cleaned map, we employ a "half-split maps" approach, where the entire dataset is divided into two segments sharing the same sky signal but having uncorrelated noise. Using cross-correlation techniques between two recovered half-split maps, we effectively reduce instrumental noise effects at the power spectrum level. As a result, we achieve precise recovery of the CMB EE and lensing B-mode power spectra. Furthermore, we also extend our pipeline to full-sky simulated data with the performance of LiteBIRD experiment. As expected, various foregrounds are cleanly removed from the foregrounds contamination observational maps,  and recovered EE and lensing B-mode power spectra exhibit excellent agreement with the true results. Finally, we discuss the dependency of our method on the foreground models. 
\end{abstract}


\section{INTRODUCTION} 
Decades of measurements of the cosmic microwave background (CMB) and its anisotropies \citep[e.g.,][]{Bennett:2003,Story:2013,Das:2014,Planck Collaboration:2016b} serve as a crucial pillar in the field of precision cosmology. Efforts are now focused on the next frontier in CMB experiments towards precise measurements of polarization anisotropies, particularly the search for the faint primordial polarization B modes. This primordial B mode originates from the primordial gravitational waves predicted by inflation, making its detection a potential direct evidence of inflation \citep{Kamionkowski:2016}. Several next-generation CMB experiments have emerged, aiming to achieve multi-frequency coverage and high sensitivity for the search of the primordial B-mode signal.  Ground-based projects like the Simons Observatory \citep{Ade:2019}, CMB-S4 \citep{Abazajian:2016}, QUIJOTE \citep{Rubino-Martin:2012,Poidevin:2018}, and AliCPT \citep{Li:2017}, as well as space-based missions like LiteBIRD \citep{Hazumi:2019,Suzuki:2018} and the Probe of Inflation and Cosmic Origins (PICO) \citep{Hanany:2019}, have been proposed or are currently being developed. However, a challenge in the analysis of CMB data lies in the extraction of the B-mode signal from observations that are contaminated by foreground radiations. The Galactic polarized radiation tends to be brighter than the primordial B-mode signal over all observational frequencies in the microwave regime \citep{Krachmalnicoff:2016,Krachmalnicoff:2018}. Consequently, the accurate separation of the foreground contaminants from the CMB observations becomes a critical task in CMB data analysis, as emphasized in studies by \cite{Betoule:2009} and \cite{Errard:2016}.

From the data analysis perspective,  the process of extracting the CMB signal from observations contaminated by foreground emissions is commonly referred to as CMB component separation. Since the various foreground components have  distinct spectral signatures and differ from that of the CMB, it is possible to reconstruct clean maps of the CMB and each foreground emission by combining observations from multiple frequencies.  Typical methods for component separation can be broadly divided into two categories: ``parametric"  and ``blind" methods. Parametric methods, such as the Commander method \citep{Eriksen:2008} and XFORECAST \citep{Errard:2016,Stompor:2016}, rely on fitting parametric models to the multi-frequency maps.  Bayesian parameter estimation or maximum-likelihood methods can then be employed to fit these parameters, achieving the goal of component separation. However, accurately modeling foreground emissions remains a complex and challenging task due to the intricate physics of foregrounds involved \citep[e.g.,][]{Draine:2013,Poh:2017}.  For future CMB B-mode detection experiments with unprecedented sensitivity, several studies \citep[e.g.,][]{Armitage-Caplan:2012,Remazeilles:2016,Hensley:2018} have reported that even slight inaccuracies in foreground modeling can lead to significant biases in the reconstruction of the CMB B-mode signal due to the larger amplitude of the Galactic polarized radiation compared to the CMB B-mode signal. This issue has also been mentioned in several next generation CMB experiments, such as the CMB-S4 project \citep{Abazajian:2016} and the CORE satellite missions \citep{Remazeilles:2018}.

On the other hand, the so-called non-parametric or blind component separation, such as the Internal Linear Combination \cite[ILC,][]{Tegmark:2004,Sudevan:2017} approach  or needlet \citep[NILC,][]{Basak:2013} or scale discretized \citep[SILC,][]{Rogers:2016}, and  Hierarchical Morphological Component Analysis  \citep[HGMCA,][]{Wagner-Carena:2020},  exploit minimal prior information on the foregrounds. Thus, non-parametric method quickly provide a foreground-cleaned CMB map, but not detailed information about the various foreground emissions.

With the notable advancements in computer science, machine learning techniques have demonstrated exceptional proficiency in the domain of image processing,  such as image recognition, restoration of noisy or blurred images, among others. Machine learning techniques have found growing applicability in the realm astrophysics as well \citep{Mehta:2019,Fluke:2020}. Notably, machine learning has been successfully employed to discern between cosmological and reionization models \citep{Schmelzle:2017,Hassan:2018}, analyze gravitational wave data \citep[e.g.,][]{George:2018,Shen:2019,Li:2020}, and reconstruct functions from cosmological observational data \citep{Wang:2020a,Wang:2021}.

In the field of CMB data processing, the application of machine learning methods for foreground subtraction in CMB temperature has been explored early on \citep{Baccigalupi:2000,Norgaard:2008}. Recently, convolutional neural network-based machine learning techniques also have shown promise in accurately extracting full-sky temperature maps of the CMB from observational data \citep{Petroff:2020, Wang:2022}, reconstructing CMB lensing \citep{Caldeira:2019,Yan:2023b}, and removing lensing effect of CMB polarization (delensing) \citep{Yan:2023a}. It should be noted that the reconstruction of CMB polarization signal poses a greater challenge compared to temperature reconstruction, as CMB polarization signal is more faint in comparison to the total Galactic polarized radiation across all observed microwave frequencies \citep{Thorne:2017}.  \cite{Yan:2023c} propose a machine-learning-based foreground-cleaning technique for CMB polarization data, called CMBFSCNN (Cosmic Microwave Background Foreground Subtraction with Convolutional Neural Networks). In \cite{Yan:2023c}' study, we first use a network model to remove polarized foreground contamination from observed data. Then, a cross-correlation technique is employed to suppress the impact of instrumental noise on the power spectrum. The results demonstrate the effectiveness of CMBFSCNN in successfully removing various foreground components from both actual observational data of the Planck experiment and simulated data. This work further applies the CMBFSCNN technique to the LiteBIRD experiment, which conducts full-sky surveys, as well as the CMB-S4 experiment, which covers the  partial sky region. The study also considers the CMB lensing effect and presents the recovery of the lensing B-mode power spectrum. Additionally, more comprehensive results regarding the CMBFSCNN technique are presented. We provide the \texttt{CMBFSCNN} code used for this analysis on \href{https://github.com/yanyepeng/CMBFSCNN}{GitHub}\footnote{https://github.com/yanyepeng/CMBFSCNN}


This paper is organized as follows. Section 2 provides a comprehensive introduction to the methodology. This includes a detail of the network structure, a concise overview of the spectral energy distribution (SED) models of diffuse Galactic foregrounds, and the data simulations.
Section 3 focuses on the application of the neural network to simulate data with the performance of CMB-S4 experiment and full-sky simulate data with the performance of LiteBIRD experiment. Section 4 is dedicated to a discussion on the utilization of the CNN method for the recovery of the CMB polarized signal. Finally, we summarize our work in Section 5.

\section{Methodology}
\subsection{Network architecture}
\label{network}
The convolutional neural network (CNN) is a class of feed-forward neural networks widely used in various fields. The convolutional layer serves as a fundamental building block of the CNN \citep{Mehta:2019}.   The convolutional layer takes feature images from the preceding layer as inputs and convolves them with multiple local spatial filters, also known as kernels.  These kernels have learnable parameters that are adjusted during training to optimize the network's performance. Then, nonlinear activation functions are applied to the outputs before passing them to subsequent layers. The configuration of a convolutional layer mainly involves three crucial hyperparameters: the number of output channels (equivalent to the number of convolutional filters), stride length, and zero padding amount. By adjusting these hyperparameters, it is possible to control the size of the output produced by each convolutional layer. Additionally, dilated convolutions have been introduced as an extension to standard convolutions for capturing more contextual information by enlarging the receptive field size \citep{Yu:2015}. By connecting multiple convolutional layers together, one can design complex network model which consists of a stack of nonlinear parameters. The network parameters can be fine-tuned  through the process of training on specific datasets, thus enabling the transformation of intricate problems into parameter optimization tasks. An example successful architecture is U-Net proposed by \cite{Ronneberger:2015}, which employs an encoder-decoder structure with additional skip connections between encoding and decoding layers to preserve small-scale information lost during downsampling operations. In the \cite{Yan:2023c}, we propose a multi-patch hierarchy network architecture based on the U-Net, and it is specifically designed for foreground removal from contaminated CMB polarization maps. This architecture draws inspiration from several related studies conducted in this domain \citep{Ronneberger:2015,Nah:2018, Tian:2020a,Tian:2020b,Syed:2021}. In this work, we adopt the same network model utilized in CMBFSCNN \citep{Yan:2023c} to remove CMB polarization contamination. Detailed information about the CNN model can be found in paper \citep{Yan:2023c}.

Once the network architecture is established, the network model parameters (weights and biases) can are optimized through a process of minimizing the loss function. The loss function serves to quantify the discrepancies between the output generated by the network and the corresponding ground truth image. In this study, our loss function consists of two components: mean absolute error (MAE), also known as L1 loss, and a loss function based on fast Fourier transform (FFT). Assuming that our training dataset comprises $S$ pairs of images denoted as $\{x_i, y_i\}_{i=1}^{S}$, where $x_i$ represents the i-th contaminated input image and $y_i$ represents its corresponding ground truth image. Let $I = f(x)$ denote the prediction produced by our network model ($f(\cdot)$). For a subsample with batch size equal to $N$, we calculate pixel-wise MAE loss as follows:
\begin{align}
	L_{\rm MAE}&=\frac{1}{N}\sum_{n=1}^{N}\lbrack \frac{1}{WH}\sum_{w=1}^{W}\sum_{h=1}^{H}(|I^{n}_{w,h}-y^{n}_{w,h}|)\rbrack,
\end{align} 
where $W$ and $H$ describe the dimensions of the images.

For the FFT loss, we define the amplitude of the FFT as follows: 
\begin{align}
	A_{\rm F}(I)& = \sqrt{Re[{\rm FFT}(I)]^2+Im[{\rm FFT}(I)]^2},
\end{align}
where $Re[\cdot]$ and $Im[\cdot]$ represent the real part and imaginary part, respectively. The notation ${\rm FFT}(\cdot)$ denotes the operation of performing  FFT.  For a subsample with the size of batch size $N$, the FFT loss function is calculated as:

\begin{align}
	L_{\rm FFT}&=\frac{1}{WN}\sum_{n=1}^{N}\lbrack \frac{1}{WH}\sum_{w=1}^{W}\sum_{h=1}^{H}(|A_{\rm F}(I^{n}_{w,h})-A_{\rm F}(y^{n}_{w,h})|)\rbrack,
\end{align}
We combine with the MAE loss and FFT loss to define our network's overall loss function as follows:
\begin{align}
	L = L_{\rm MAE} + \beta L_{\rm FFT},
\end{align}
where $\beta$ is a hyperparameter and we set it as $1$ according to our experience test.

By incorporating the defined loss function, our target during the training process is twofold. The network model aims not only to minimize discrepancies between predicted and ground truth maps at the pixel level but also to ensure closeness in terms of their respective amplitudes in the fast Fourier transform (FFT) domain. The inclusion of FFT amplitudes in the loss function stems from experiential testing, where it has been observed that utilizing an FFT loss function can slightly enhance recovery of the CMB power spectrum.

\subsection{Foreground Parametrization}
\label{data_foreground}

The Convolutional Neural Network (CNN) proposed in this study is a supervised machine learning algorithm, which requires the utilization of a training dataset comprised of known ground truth values. In our case, the training samples are obtained from simulated data generated using the publicly available Python Sky Model (\texttt{PySM}\footnote{https://github.com/bthorne93/PySM\_public}) package \citep{Thorne:2017}. The \texttt{PySM} package enables simulation of full-sky maps including Galactic emission in both intensity and polarization at microwave frequencies. We focus specifically on three polarized foreground sources: synchrotron radiation, thermal dust emission, and Anomalous Microwave Emission (AME).

\subsubsection{Synchrotron}
\label{Synchrotron}

Synchrotron radiation arises from the acceleration of relativistic cosmic ray electrons by Galactic magnetic fields. Below frequencies of approximately $\sim 50$ GHz, it constitutes the dominant source of polarized foregrounds \citep{Kogut:2007}. The spectral energy distribution (SED) of synchrotron emission is commonly described by a power law function. In this study, we adopt a general model to represent synchrotron polarized emission, which can be expressed as:

\begin{align}
	Q_s(\hat{n},\nu)&=A_{Q,s_{\nu_{0}}}(\hat{n})\left(\frac{\nu}{\nu_{0}}\right)^{\beta_s(\hat{n})},\\
	U_s(\hat{n},\nu)&=A_{U,s_{\nu_{0}}}(\hat{n})\left(\frac{\nu}{\nu_{0}}\right)^{\beta_s(\hat{n})}.
\end{align} 
$Q_s(\hat{n},\nu)$ and $U_s(\hat{n},\nu)$ represent the synchrotron Stokes polarization components. The amplitudes of these quantities at the pivot frequency $\nu_0$ are denoted as  $A_{Q,s_{\nu_{0}}}$ and $A_{U,s_{\nu_{0}}}$. The synchrotron spectral index is represented by $\beta_s$. It should be noted that all these parameters exhibit spatial dependence due to their expected variability across different sky directions ($\hat{n}$).

The simulation of Galactic synchrotron polarized radiation is performed by extrapolating template maps based on a parametric model derived from the \texttt{PySM} s1 model. The polarization template maps are constructed using WMAP 9-yr 23 GHz Q and U maps \citep{Bennett:2013}, respectively.  The spectral index map is obtained from "Model 4" presented in \cite{Miville:2008}, which combines data from Haslam and WMAP 23-GHz polarization observations along with a Galactic magnetic field model. This spectral index map exhibits spatial variability, characterized by a mean value of approximately $-3$ with a error of around $0.06$.

\subsubsection{Thermal dust}
The emission of thermal dust radiation originates from interstellar dust grains, which are heated through absorption in the optical and subsequently cooled by emitting in the far-infrared regime. At frequencies above approximately $\sim 70$ GHz, it constitutes the primary source of polarized foregrounds. The SED of thermal dust is characterized by a modified blackbody emission due to opacity effects. As a result, the observed polarized spectrum is commonly described using a modified black-body model:
\begin{align}
	Q_d(\hat{n},\nu)&=A_{Q,d_{\nu_{0}}}(\hat{n})\left(\frac{\nu}{\nu_{0}}\right)^{\beta_d(\hat{n})}B(\nu,T_d(\hat{n})),\\
	U_d(\hat{n},\nu)&=A_{U,d_{\nu_{0}}}(\hat{n})\left(\frac{\nu}{\nu_{0}}\right)^{\beta_d(\hat{n})}B(\nu,T_d(\hat{n})).
\end{align} 
$Q_d(\hat{n},\nu)$ and $U_d(\hat{n},\nu)$ represent its Stokes polarization components. The amplitudes of these quantities at the pivot frequency $\nu_0$ are denoted as  $A_{Q,d_{\nu_{0}}}$ and $A_{U,d_{\nu_{0}}}$. The spectral index is represented by $\beta_d$.  Additionally, we use the function $B(\nu,T_d(\hat{n}))$, which corresponds to a standard black body spectrum with temperature  $T_d(\hat{n}) \approx 20$ K.

To simulate polarized dust maps, we utilize the template maps, spectral index map, and temperature ($T_d(\hat{n})$) map derived from the \texttt{PySM} d1 model. This particular model employs polarization template maps at 353 GHz, which are obtained through analysis of Planck data using the \texttt{COMMANDER} code \citep{Planck Collaboration:2016a}.  The spatial distribution of both $\beta_d$ (the spectral index) and $T_d(\hat{n})$ (the temperature) exhibit variability across different sky directions. Their respective mean values are approximately $\beta_d\approx 1.54\pm0.03$ and $T_d(\hat{n})\approx 20.9\pm2.2$.

\subsubsection{Anomalous microwave emission}

Anomalous Microwave Emission (AME) has been observed by radio/microwave instruments within the frequency range of approximately $\approx 10-60$GHz \citep{Kogut:1996,Leitch:1997,Murphy:2010}. Despite its detection, the precise mechanism responsible for AME remains uncertain.  A notable and promising candidate model is based on electric dipole radiation emitted from small spinning dust grains \citep{Erickson:1957,Draine:1998,Draine:1999,Ysard:2010},  which is adopted as the working model for AME in this study. This particular model attributes the emission to rotational motion of a dust grain possessing an electric or magnetic dipole moment. It has been established that AME exhibits a low level of polarization. For instance, \cite{Dickinson:2011} constrained the polarization fraction of AME to be less than $2.6$\% using WMAP 7-year data. Similarly, \cite{Genova-Santos:2017} utilized QUIJOTE's data at a frequency of $17$ GHz to set upper limits on the polarization fraction of AME at $0.39$\%, which further decreased to $0.22$\% when combined with WMAP's data at $41$ GHz. Although AME demonstrates a relatively low degree of polarization, it still represents a potentially significant foreground component for future sensitive CMB experiments aiming to detect CMB B-modes \citep{Remazeilles:2016}.

In this study, we adopt the AME model from the \texttt{PySM} a2 model. Within this framework, the AME intensity is determined using the {SPDUST2} code \citep{Ali-Haimoud:2009,Silsbee:2011}, which relies on Planck templates derived from the \texttt{COMMANDER} parametric fit to the Planck data \citep{Planck Collaboration:2016a}. The mathematical expression for the AME intensity can be formulated as follows:
\begin{equation}
	\begin{split}
		I_a(\hat{n},\nu)=&A_{T,\nu_{0,1}}(\hat{n})\epsilon(\nu,\nu_{0,1},\nu_{p,1}(\hat{n}),\nu_{p_0})\\
		&+A_{T,\nu_{0,2}}(\hat{n})\epsilon(\nu,\nu_{0,2},\nu_{p,2}(\hat{n}),\nu_{p_0}).
	\end{split}
\end{equation}
The AME polarization model makes use of the dust polarization angle, denoted as $\gamma_{\rm d}$, to generate a template map. This angles is calculated using the Planck \texttt{COMMANDER} 2015 thermal dust Q and U maps at a frequency of 353 GHz. The expression for the AME polarization can be represented as:
\begin{align}
	Q_a(\hat{n},\nu)&=fI_{a}{\rm cos}(2\gamma_{353}),\\
	U_a(\hat{n},\nu)&=fI_{a}{\rm sin}(2\gamma_{353}),
\end{align} 
where $f$ is the polarization fraction, which is set at a global value of 2\% in this work. 

\subsection{Data Simulations}
\label{simulations}

\begin{figure*}
	\centering
	\includegraphics[width=1.0\hsize]{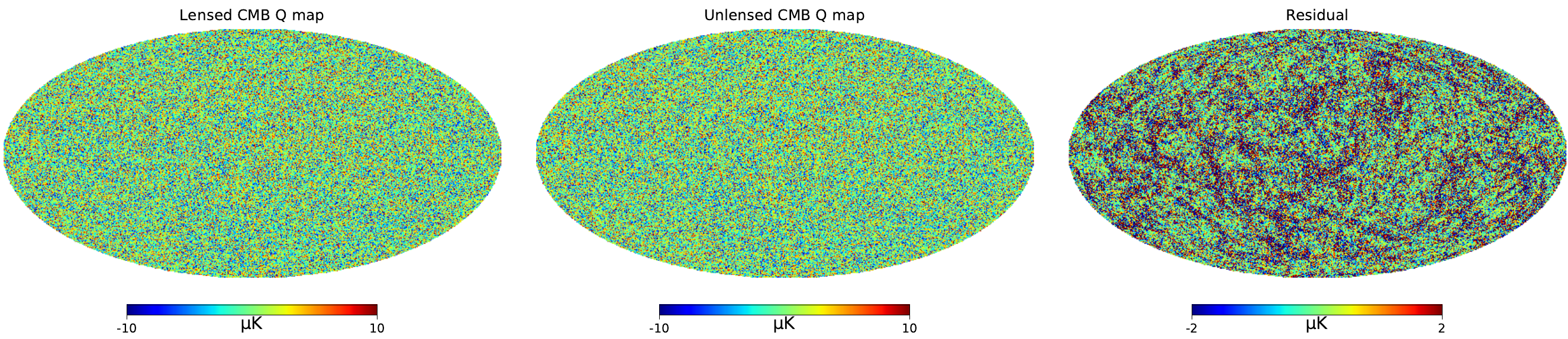}
	\caption{Lensed and unlensed CMB Q maps and their residual map.}  
	\label{figure_lensed_cmb}
\end{figure*}

For CMB simulation, we first use \texttt{CAMB}\footnote{https://github.com/cmbant/CAMB} software package  to calculate the CMB power spectra and lensing power spectrum within $\Lambda$  cold dark matter cosmological framework ($\Lambda$CDM framework).  The cosmological parameters of $\Lambda$CDM framework have $H_0$, $\Omega_bh^2$, $\Omega_ch^2$, $\tau$, $A_s$, $n_s$, and their best-fit value and standard deviation ($1\sigma$) is obtained from the Planck 2015 data \citep{Planck Collaboration:2016c}. Next, we generate  CMB maps and lensing potential maps using the publicly available software package called \texttt{Healpy} \footnote{\url{https://github.com/healpy/healpy}}, which is a Python wrapper for HEALPix\footnote{ \url{https://healpix.sourceforge.io/downloads.php}} developed by \cite{Gorski:2005}. These maps have a pixel resolution defined by $N_{\rm side}$ = 512.  The lensed CMB maps are created by lensing the primordial CMB map with the lensing potential maps utilizing the \texttt{TAYLENS} code provided in \cite{Naess:2013}. Figure \ref{figure_lensed_cmb} illustrates an example of both lensed and unlensed CMB Q maps. Notably, due to gravitational lensing effects, significant information is retained in residual maps. Foreground components are generated via implementation of the foreground model introduced in Section \ref{data_foreground}.

\begin{table*}
	\begin{center}
		\centering
		\caption{Frequencies and instrumental specifications of the LiteBIRD satellite and the CMB-S4 experiment \citep{Krachmalnicoff:2022,Zegeye:2023}.}\label{table_1}
		\begin{tabular}{ |c|c|c|c| }
			\hline
			Experiment & Frequency & White noise level in polarization  & FWHM  \\
			-& (GHz) & ($\mu$K-arcmin) & (arcmin)\\
			\hline
			CMB-S4 & 30.0 & 3.53 & 72.8 \\
			& 40.0 & 4.46 & 72.8 \\
			& 85.0 & 0.88 & 25.5 \\
			& 95.0 & 0.78 & 22.7 \\
			& 145.0 & 1.23 & 25.5 \\
			& 155.0 & 1.34 & 22.7 \\
			& 220.0 & 3.48 & 13.0 \\
			& 270.0 & 5.97 & 13.0 \\
			
			\hline\hline
			LiteBIRD  
			& 50.0 & 32.78 & 56.8 \\
			& 78.0 & 18.59 & 36.9 \\
			& 100.0 & 12.93 & 30.2 \\
			& 119.0 & 9.79 & 26.3 \\  
			& 140.0 & 9.55 & 23.7 \\
			& 166.0 & 5.81 & 28.9 \\ 
			& 195.0 & 7.12 & 28.0 \\ 
			& 235.0 & 15.16 & 24.7 \\
			& 280.0 & 17.98 & 22.5 \\
			& 337.0 & 24.99 & 20.9 \\
			\hline
		\end{tabular}
	\end{center}
\end{table*}

\begin{figure*}
	\centering
	\includegraphics[width=1.0\hsize]{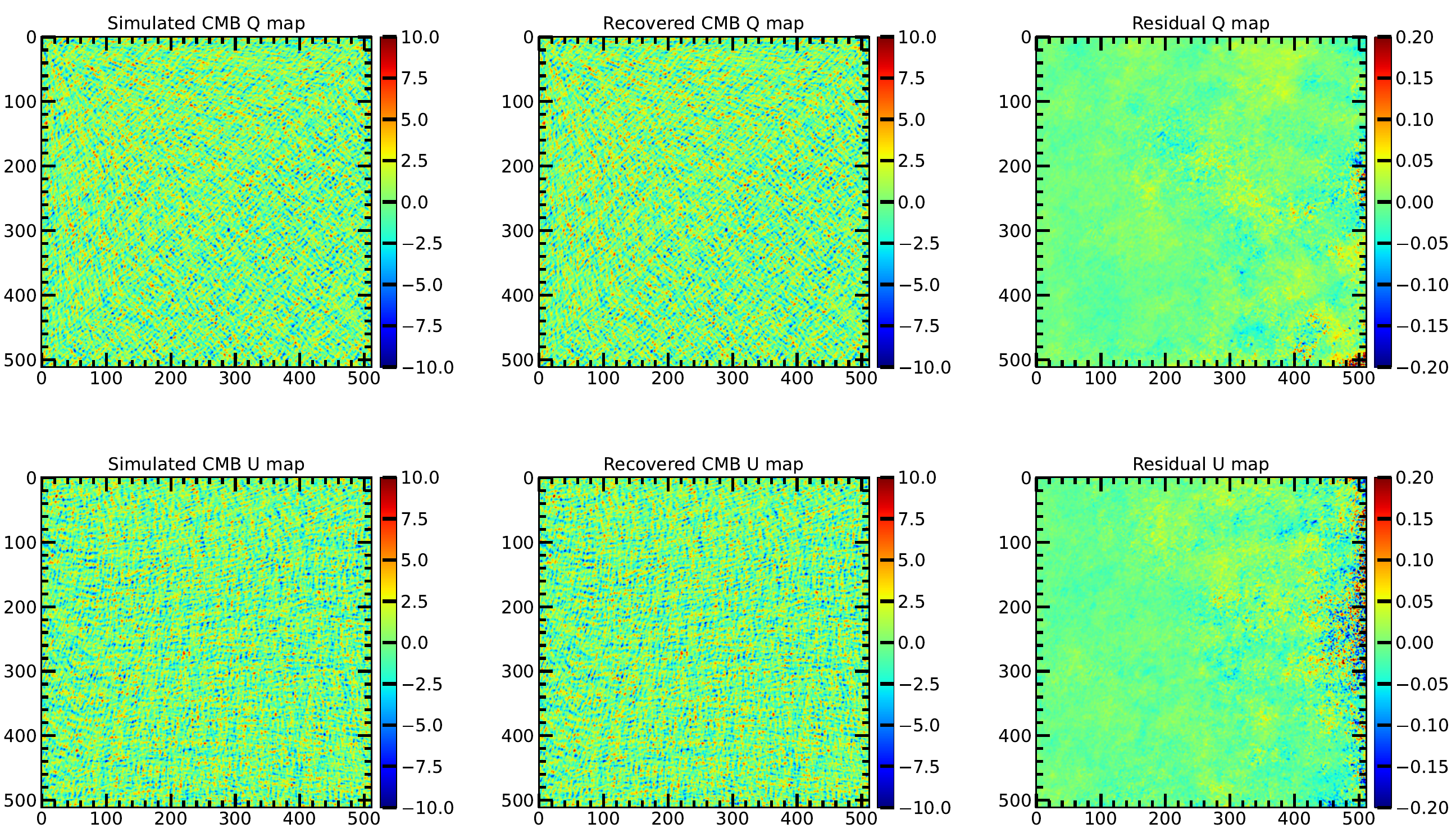}
	\caption{The recovered noisy CMB polarization maps for the partial-sky observation data with the performance of CMB-S4 experiment. The upper panels depict the recovery of the Q map, while the lower panels  display the recovery of the U map. The simulated maps utilized in this analysis comprise the beam-convolved CMB map, plus a noise map at a frequency of 220 GHz.The recovered maps correspond to the noisy CMB maps reconstructed using the {\tt CMBFSCNN} algorithm. }  
	\label{fig:map_2sigma}
\end{figure*}

\begin{figure*}
	\centering
	\includegraphics[width=1.0\hsize]{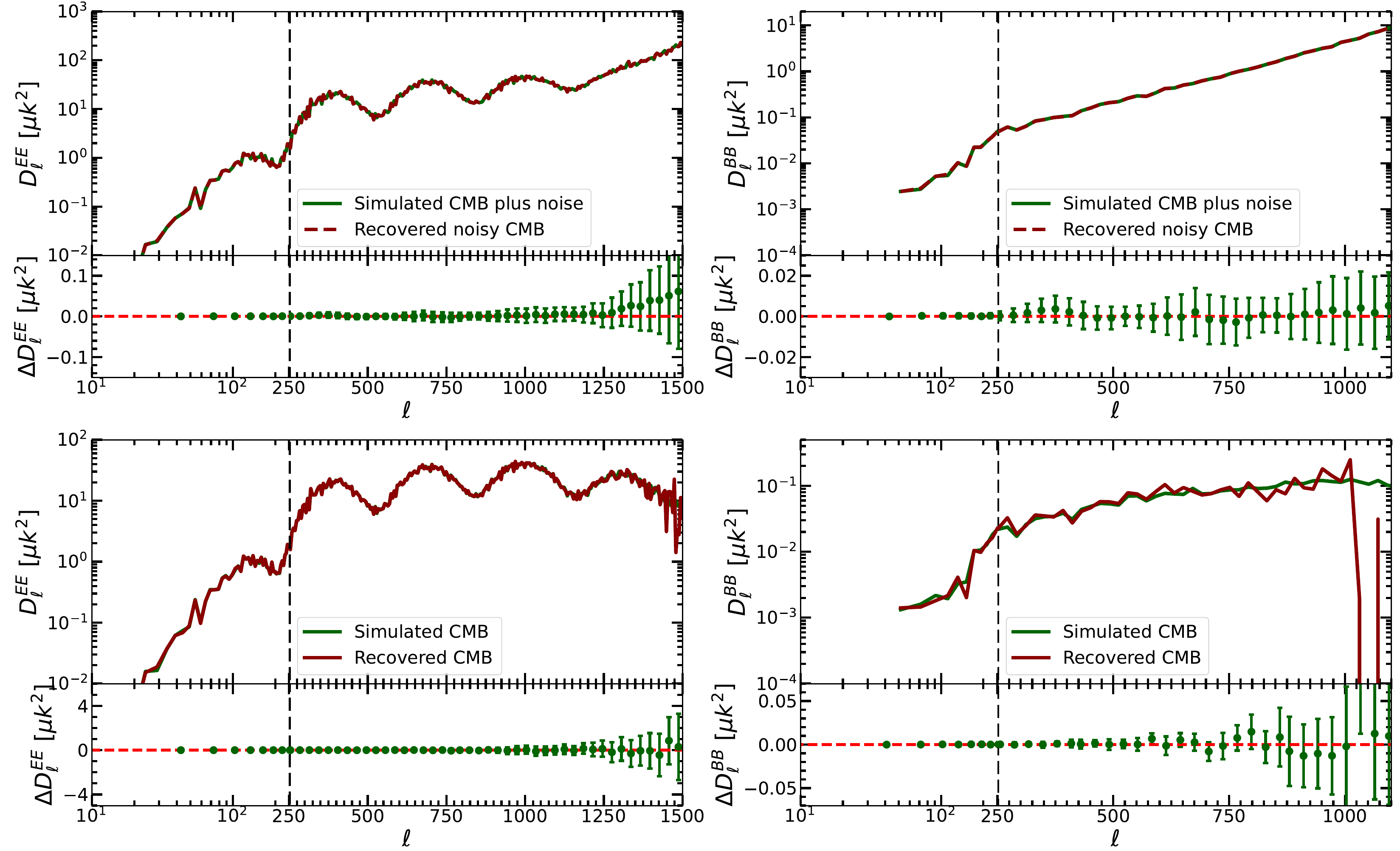}
	\caption{ Recovered EE and lensing B-model power spectra for the partial-sky observation data with the performance of CMB-S4 experiment. The two upper panels present the EE and BB power spectra from the recovered noisy CMB maps.  The two lower panels show the final EE and BB power spectra, after the denoising procedure. Additionally, the deviations between the recovered power spectra and the input fiducial power spectra are also illustrated. The length of each $\ell$ bin is set to be 5 for the EE power spectrum and 20 for the BB power spectrum here.  We choose the logarithm scales  for $\ell$ within the angle scale range of $\ell \le250$, and the linear scales within the angle scale range of $\ell>250$. }  
	\label{figure_s4_cross_QUEB}
\end{figure*}

Our CNN model is designed to learn the mapping between contaminated polarization maps and foreground-cleaned CMB polarization maps. In the context of machine learning, the ability of a trained neural network to accurately predict unseen data is referred to as generalization. We also hope that the network has sufficient generalization capability to handle the real data. To achieve this, we generate training data that closely resembles real-world scenarios. However, the foreground emission is too complex \citep{Finkbeiner:1999,Kogut:2007,Kogut:2012} to be parameterized, and this is reflected as the complicated spatial variation of the amplitude and spectral index in the power-law model of synchrotron, the modified blackbody model of thermal dust, and the model of AME. Consequently, during training data generation processes, we introduce manual uncertainties into these parameters. Specifically, for CMB simulations, cosmological parameters are treated as independent Gaussian random variables using mean values and $1\sigma$ standard deviations derived from Planck 2015 results. On the other hand, for foreground realizations involving amplitude template map ($A$) and spectral index map ($\beta$), each pixel value is multiplied by a randomly generated number with an average value of 1 and a standard deviation of $0.1$ ($0.05$ for a spectral index).

In this work, our target is to use the network model to remove the foregrounds from the multi-frequency observational maps. The inputs to the network model are the beam convolved observational maps at multiple frequencies, which contain the CMB, foregrounds, and instrumental noise. The desired output of network is the beam-convolved CMB maps (with lensing effect) plus noise maps. This implies that our network is designed to only remove the foregrounds while preserving beam convolved CMB with lensing effect and noise. The simulated CMB and foreground maps are represented as one-dimensional arrays using the RING numbering scheme of HEALPix. However, since the neural network utilized in this study requires two-dimensional data for both inputs and outputs, it is necessary to convert these one-dimensional arrays into two-dimensional arrays before inputting them into the network. We initially divide each input map into 12 patches according to the NESTED ordering scheme. Subsequently, each data patch is directly filled with an $N_{\mathrm{side}} \times N_{\mathrm{side}}$ square grid before combining these grids into a single 2D map representation. To obtain output maps in HEALPix format, an inverse process can be applied on the outputs generated by the network. For more detailed information regarding our methodology and implementation, please refer to \citet{Wang:2022}. Finally, angular power spectra $C_\ell$ are computed using \texttt{NaMaster}\footnote{\url{https://github.com/LSSTDESC/NaMaster}} software package \citep{Alonso:2019}.

\section{APPLICATION TO CMB EXPERIMENTS}
\label{Noise_Case}

\subsection{Application to CMB-S4 experiment}
\label{CMB_S4_Experiment}

In this section, we apply our method to a set of simulated data that is representative of the performance characteristics of the CMB-S4 experiment. Specifically, we generate 1000 beam convolved emission maps and 300 white noise maps, encompassing eight distinct frequency bands. The chosen ${\rm N_{side}}$ parameter value is set at 512. We provide a summary of the frequency bands employed and instrumental properties specific to the CMB-S4 experiment in Table \ref{table_1}.  Here, we assume that the instrument noise is Gaussian and white.  For each individual frequency band, these 300 white noise maps are randomly added to 1000 beam convolved emission maps, thereby creating a training dataset consisting of 1000 observed maps across all eight frequency bands. To construct an independent test set for evaluation purposes, we generate an additional set of 300 sky emission maps and 300 noise maps using distinct random seeds and parameter values. Notably, it is important to highlight that the simulation process for the CMB maps incorporates considerations for lensing effects.

The inputs to our neural network model comprise beam-convolved observational $Q$ or $U$ maps obtained from all eight frequency bands, encompassing lensed CMB signals, foreground emissions, and instrumental noise. The desired output from the network is a beam-convolved lensed CMB map ($Q$ or $U$) with the noise and beam at 220 GHz. To optimize the performance of our model, we employ the Adam optimizer algorithm as proposed by \cite{Kingma:2014}, initializing the learning rate at $0.01$ and progressively reducing it during iterations until reaching a value of $10^{-6}$. The training process encompasses approximately 30,000 iterations using a batch size of 12 and is executed on two NVIDIA Quadro GV100 GPUs. On average, the training process for a single network model requires approximately 14 hours to complete.

In this study, we select a patch of the sky that is one-twelfth the size of the entire sky, containing $512\times512$ pixels, as the training dataset for our neural network. Subsequently, we apply the trained network to a sample from the testing set. The resulting reconstructed CMB $Q$ and $U$ maps are presented in Figure \ref{fig:map_2sigma}. Upon examination of the residual map, it can be observed that the reconstructed polarization maps closely resemble the true simulated maps. To quantitatively evaluate the performance of our network, we utilize the mean absolute difference (MAD) as a metric, employing the following general formula:
\begin{equation}
	\begin{split}
		\sigma_{\rm MAD} &= \frac{1}{N}\sum_{i}^{N}\left|X_i - Y_i\right|,
	\end{split}
\end{equation}
where $N$ is the number of pixels, $X$ and $Y$ represent the predicted and real sky maps. The MAD between the recovered CMB $Q$ and $U$ maps and the target $Q$ and $U$ maps is computed. For the test set, the MAD values for the recovered $Q$ map and $U$ map are determined to be $0.016 \pm 0.008\ \mu$K and $0.021 \pm 0.002\ \mu$K, respectively. Similarly, the average MAD values for the training set are found to be $0.015 \pm 0.003\ \mu$K for the $Q$ map and $0.020 \pm 0.002\ \mu$K for the $U$ map, exhibiting consistency with the results obtained from the test set. It is worth noting that the performance of the recovered maps in the right region, as depicted in panel (a) or (b) of Figure \ref{fig:map_2sigma}, is slightly worse than that for other areas. This discrepancy can be attributed to the proximity of the right region to the galactic plane, resulting in significant foreground contamination.

%

As shown in the upper panels of Figure \ref{figure_s4_cross_QUEB}, we derive the recovered noisy CMB $EE$ and $BB$ power spectra from the recovered $Q$ and $U$ maps. The recovered noisy $EE$ and $BB$ power spectra closely resemble the target spectra, indicating the ability of our network to effectively remove foreground contamination at the power spectrum level. It should be noted that the recovered CMB spectra show a rapid increase for scales $\ell>1000$, primarily due to the beam effects and instrumental noise. In the Figure \ref{figure_s4_cross_QUEB}, the deviation of the power spectra ($\Delta^{EE}_{\ell}$ or $\Delta^{BB}_{\ell}$ ) is the average deviation  across the test set, utilizing a bin size of $\ell=30$.  The error bars ($\sigma^{b}$) are calculated as follows: we first compute the standard deviation ($\sigma_{\ell}$) of the power spectra on the test set, 
\begin{equation}
\begin{split}
\sigma_{\ell}=\sqrt{\frac{\sum_{i=1}^{N}(D^{i,{\rm predicted}}_{\ell}-D^{i,{\rm target}}_{\ell})^{2}}{N}}
\end{split}
\end{equation}
where, $D^{i,{\rm predicted}}_{\ell}$ and $D^{i,{\rm target}}_{\ell}$ represent predicted power spectrum by our method and true power spectrum, respectively. Symbol $N$ is the number of samples in the test set. Subsequently, employing $\ell=30$ as the binning criterion, the standard deviation within a bin $\sigma^{b}_{\ell}$ is denoted as
\begin{equation}
\begin{split}
\sigma^{b} = \sqrt{\sum_{\ell=b*30}^{(b+1)+30} \frac{1}{\sigma_{\ell}^2}},
\end{split}
\end{equation}
here,  $b=0,1,..$ and represents index of the bin.

In our proposed methodology, the output CMB maps from the neural network retain the noise originating from the instrument. To mitigate the impact of instrumental noise at the power spectrum level, we draw inspiration from the approach employed by \citet{Krachmalnicoff:2022}. Specifically, according to CMB experiments scanning strategies, in a complete scan of the sky, we can divide the entire observational data into two parts based on the order of observation time. Thus, we can partition the entire data into two "half-split (HS) maps". These two HS maps share the same sky signal because the observed sky coverage remains unchanged, but two HS maps possess uncorrelated instrument noise because the observations are made at different time. Subsequently, we calculate the cross-correlation power spectra of two HS maps. As a result of the uncorrelated instrumental noise, the noise effects become nearly negligible in the cross-correlation power spectra, while the signal remains intact. Consequently, the cross-correlation between two HS maps provide an estimation of the signal power spectrum. It is important to note that the noise in the HS maps is enhanced by a factor of $\sqrt{2}$ relative to the sensitivity values provided in Table \ref{table_1}.

After obtaining the foreground-cleaned HS maps through the neural network's output, we employ the cross-spectra between the recovered noisy HS maps to obtain the CMB power spectra. The outcomes of this procedure are depicted in the lower panels of Figure \ref{figure_s4_cross_QUEB}. Notably, the recovered EE angular power spectra exhibit a remarkable consistency with the simulated spectra. However, by inspecting the discrepancy, $\Delta D_{\ell,{\rm CNN}} =D_{\ell,{\rm recovered}} - D_{\ell, {\rm true}}$, between the recovered and fiducial power spectra, it becomes evident that the error in the recovered $EE$ power spectra progressively increases for $l>1200$ due to the influence of beam effects. Additionally, we employ the coefficient of determination,  $R^2 = 1-\sigma_{\rm CNN}^2/\sigma^2$, as a metric to assess the recovered performance on the power spectrum. Here, $\sigma_{\rm CNN}^2$ is computed as
\begin{equation}
	\begin{split}\label{MS}
		\sigma_{\rm CNN}^2 &= \frac{1}{N}\sum_{i}^{N}\left(X_i - Y_i\right)^2,
	\end{split}
\end{equation}
where $N$, $X$ and $Y$ are the maximum multipoles ($\ell_{\rm max}=1500$), $D_{\ell,{\rm recovered}}$ and $D_{\ell, {\rm true}}$, respectively. $\sigma^2$ represents the variance of the true CMB power spectrum, which corresponds to the power spectrum derived from the input fiducial map. $R^2=1$ indicates an exact match between the recovered power spectrum and the fiducial power spectrum. Conversely, a lower $R^2$ value (its minimum value is 0) signifies poorer fitting performance. 

In our practical implementation, we evaluate the effectiveness of the recovery signal process by quantifying the $R^2$ value for the recovered power spectrum. Specifically, for the recovered EE power spectrum, we obtain $R^2 = 0.997 \pm 0.012$ (68\% C.L.) across all scales and $R^2 = 0.9998 \pm 0.0001$ (68\% C.L.) for angular scales $\ell<1200$. These results demonstrate a excellent agreement between the recovered spectrum and the fiducial spectrum.

\begin{figure*}
	\centering
	\includegraphics[width=1.0\hsize]{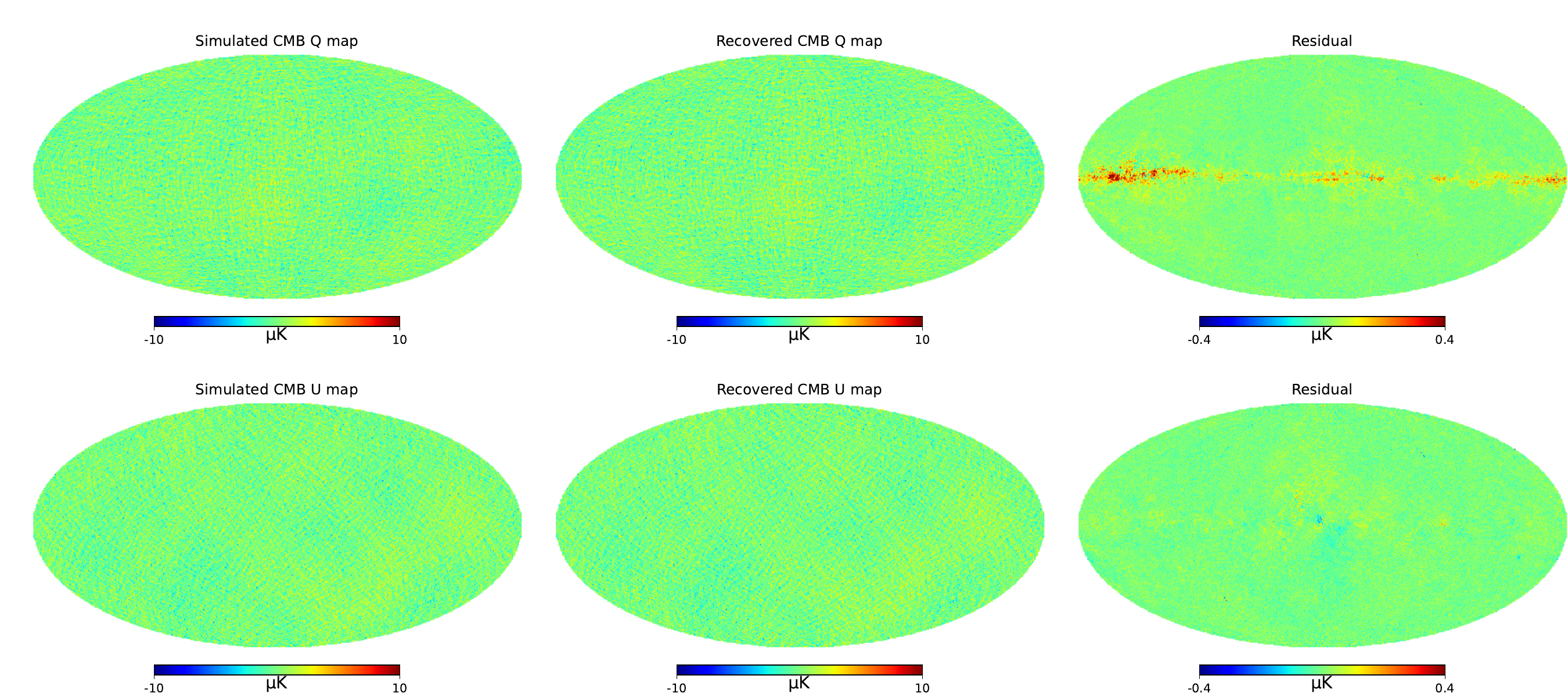}
	\caption{The recovered noisy CMB polarization maps for full-sky observation data with the performance of LiteBIRD experiment. The upper panels depict the recovery of the Q map, while the lower panels display the recovery of the U map. The simulated CMB maps utilized in this analysis comprise the beam-convolved CMB map plus an ILC noise map. The recovered CMB maps correspond to the noisy CMB maps reconstructed using the {\tt CMBFSCNN} algorithm. Additionally, the corresponding residual maps are also represented.}
	\label{fig:map_2sigma_lite}
\end{figure*}
\begin{figure*}
	\centering
	\includegraphics[width=0.9\hsize]{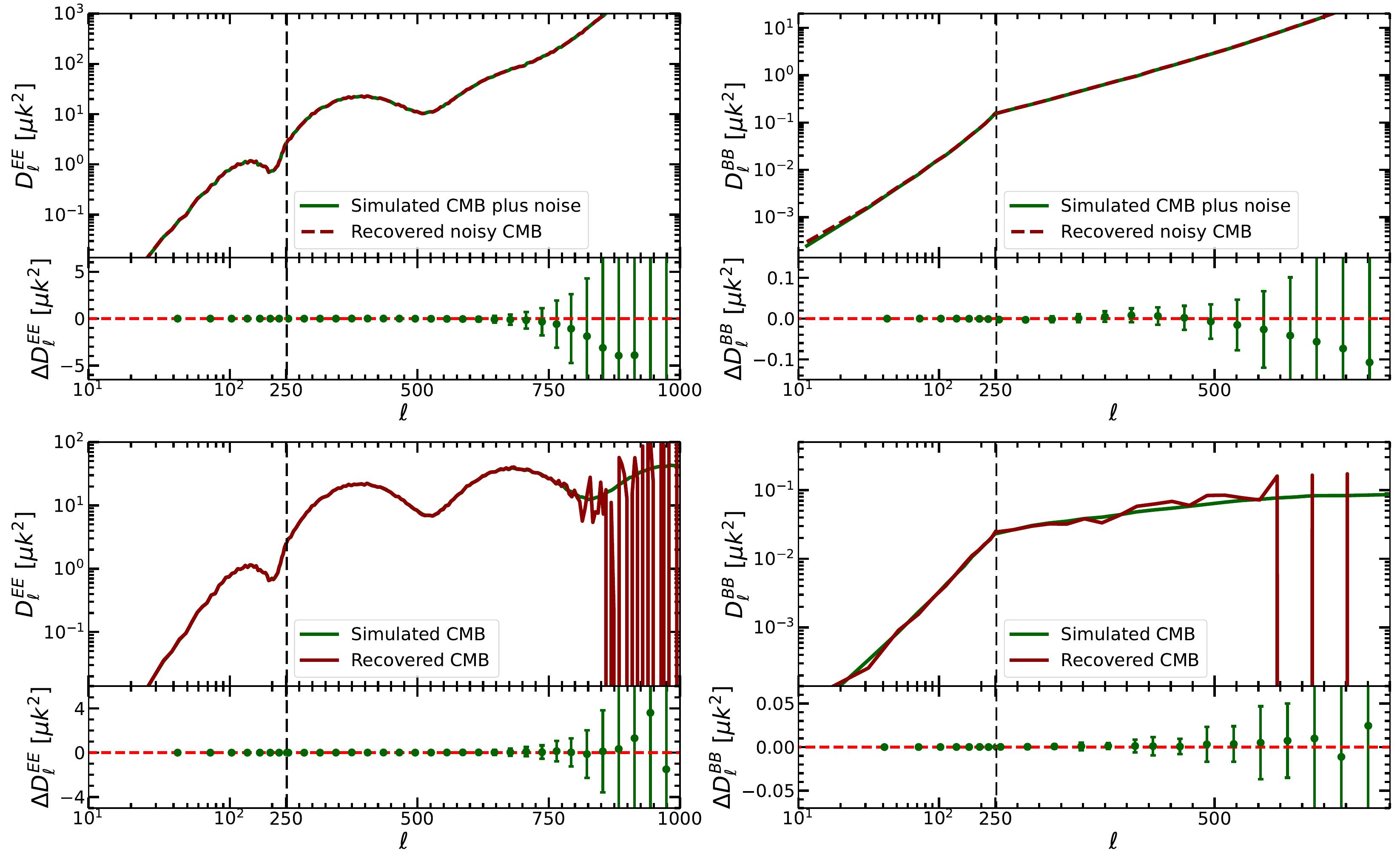}
	\caption{The recovery of EE and lensing B-model power spectra for full-sky observation data with the performance of LiteBIRD experiment. The two upper panels show the CMB EE and lensing BB power spectra of the recovered noisy CMB maps. The two lower panels show the final EE and BB power spectra, after the denoising step. Additionally, the deviations between the recovered power spectra and the input fiducial power spectra are also presented.}  
	\label{figure_s4_cross_QUEB_lite}
\end{figure*}

In the lower right panel of Figure \ref{figure_s4_cross_QUEB}, we present the recovery of the lensing $BB$ power spectrum.  It is evident that the lensing B-mode power spectrum can be successfully recovered for angular scales $\ell<800$. Subsequently, we evaluate the recovery performance by calculating the $R^2$ value for the recovered lensing B-mode power spectrum. For angular scale $\ell<800$, we find $R^2 = 0.95 \pm 0.028$ (68\% C.L.), while for $\ell<600$, we obtain $R^2 = 0.98 \pm 0.015$ (68\% C.L.), indicating a good agreement between the recovered $BB$ spectrum and the fiducial spectrum. However, we observe a limitation in obtaining the lensing B-mode power spectra for angular scales $\ell>800$, attributed to the presence of noise.

\subsection{Application to LiteBIRD experiment}
\label{lite_exp}
Considering the successful removal of foregrounds achieved by our network in the performance of CMB-S4 experiment, we extend the application of this pipeline to full-sky simulated data corresponding to the performance of the LiteBIRD experiment. Similar to the data simulation procedure described in Section \ref{CMB_S4_Experiment} for the CMB-S4 experiment, we simulate 1000 observed emission maps that emulate the performance of the LiteBIRD experiment. The frequency bands and instrumental characteristics of the LiteBIRD experiment are summarized in Table \ref{table_1}. The training set comprises 1000 observed maps acquired at ten frequency bands, while the test set comprises 300 observed maps.

As indicated in Table \ref{table_1}, the LiteBIRD experiment exhibits a higher instrumental noise level compared to the CMB-S4 experiment. In the process of using CNN for CMB component separation, it is impossible to completely eliminate the instrumental noise as shown in Section \ref{map_denoise}. In order to minimize the contamination of noise on the recovered CMB signal as much as possible, we calculate the minimized variance of noise from multiple observed frequency bands. Thus, we can obtain noise with minimized variance, expecting it to be lower than the noise level in each individual frequency map due to the accumulation of information. Ultimately, we add the noise with minimized variance to the CMB sky map as the desired output of the CNN model, which can minimize the contamination of noise on the CMB sky map output by the CNN model.

Internal Linear Combination (ILC) is one method for computing the minimized variance, so we adopt the ILC method to calculate the noise with minimized variance. We employ the ILC method to obtain a weighted sum of the noise maps corresponding to the ten frequency bands of LiteBIRD experiment. Specifically, we 
use the polarization ILC method \citep{Tegmark:2004, Kim:2009, Fern:2016, Zhang:2022} to compute the target noise map with minimized variance. This method allows us to express the processed map as follows:
\begin{equation}
	\label{eq:ilc_QU}
	\hat{Q}(p)\pm i\hat{U}(p) = \sum_{f} \left(\omega^R_f\pm i\omega^I_f\right)\left(Q_f(p)\pm i U_f(p)\right)~.
\end{equation}
As an un-biased estimator, corresponding linear weights should satisfy the following conditions, reads:
\begin{align}
	\sum_f\omega^R_f=1, \sum_f\omega^I_f=0~,
\end{align}
and can be obtained by minimizing the variance of ${|\hat{Q}+i\hat{U}|}^2$. $p$ and $f$ stand for the pixel index and the frequency channel. We perform the polarization ILC on each training set consisting of the lensed CMB, the foreground emission from the \texttt{PySM} package, and noise, to get the corresponding weight coefficients $\omega^R$ and $\omega^I$. By applying these coefficients, we obtain the target noise map as a weighted summation of the noise maps corresponding to the ten frequency channels. The inputs to the network model consist of full-sky observational maps ($Q$ or $U$) acquired at ten distinct frequencies.  The desired output of the network is a full-sky CMB map ($Q$ or $U$) convolved with the beam at 166 GHz, plus an ILC noise map derived from a weighted summation of the noise maps of the ten frequency bands.  It is important to note that the training data of the network use the HS maps, which implies that the standard deviation of the noise for the HS maps is amplified by a factor of $\sqrt{2}$.

\begin{figure*}
	\centering
	\subfigure[Case 1]{
		\includegraphics[width=17cm]{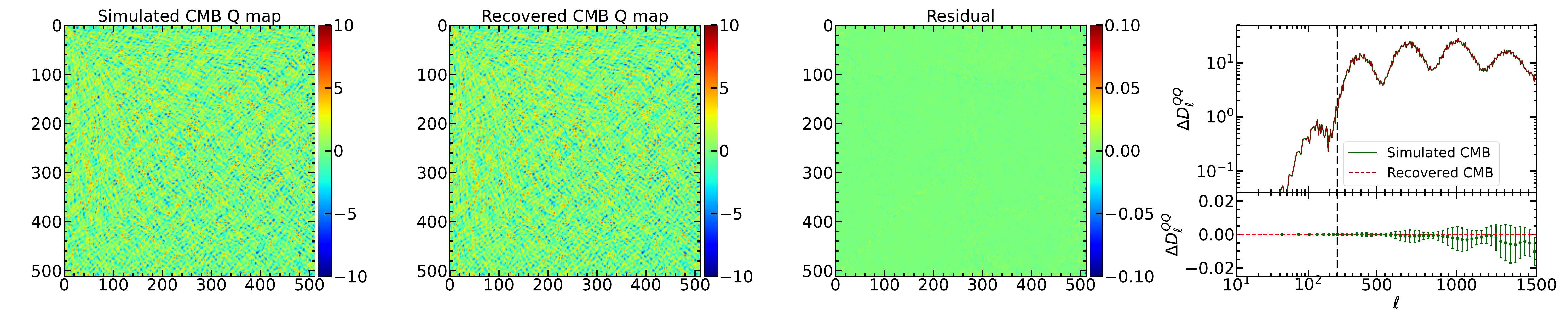}
	}
	\vspace{-1mm}
	\subfigure[Case 2]{
		\includegraphics[width=17cm]{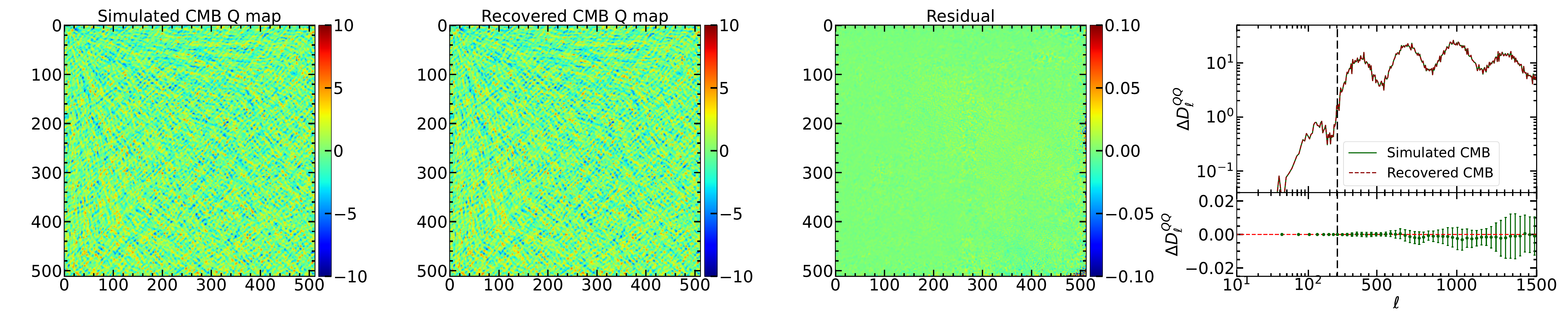}
	}
	\quad
	\subfigure[Case 3]{
		\includegraphics[width=17cm]{CMB_S4_QQ_one_A10B05.pdf}
	}
	\vspace{-1mm}
	\subfigure[Case 4]{
		\includegraphics[width=17cm]{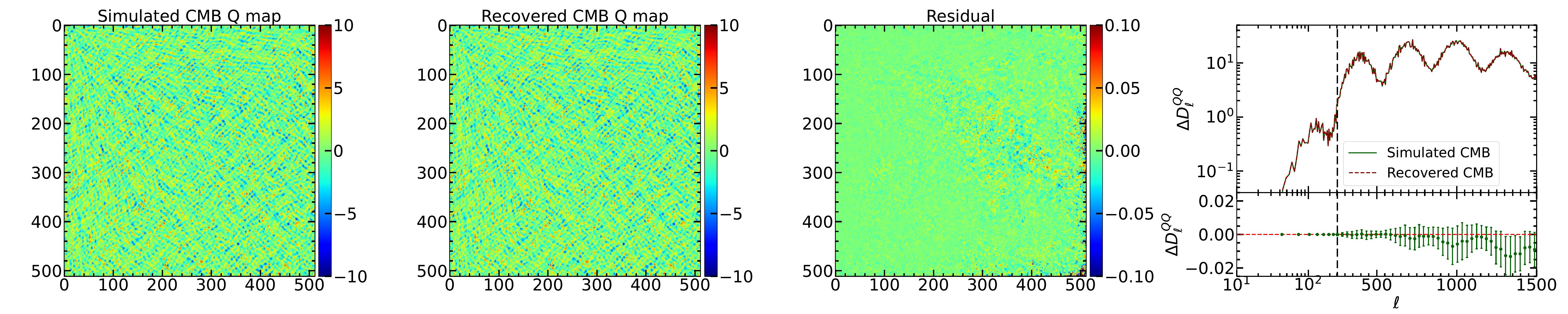}
	}
	\subfigure[Case 5]{
		\includegraphics[width=17cm]{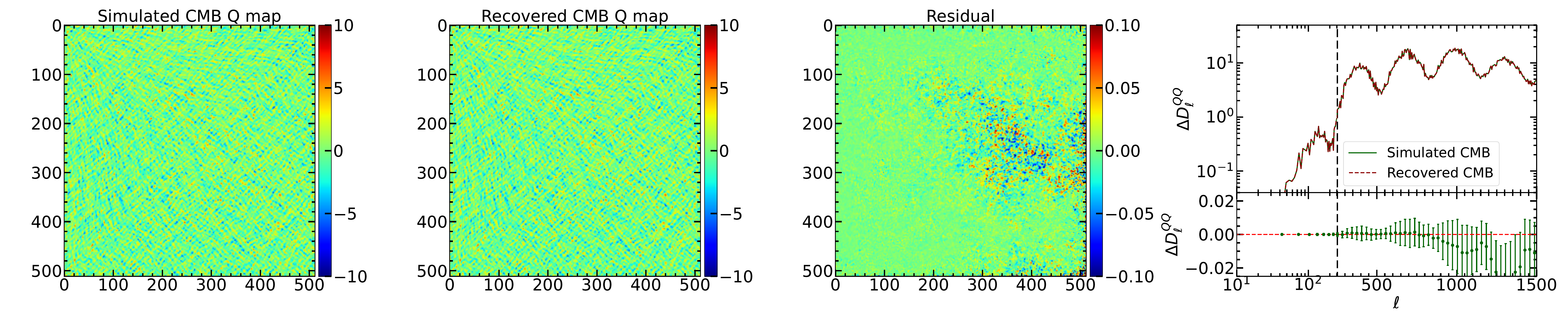}
	}
	\caption{Recovered CMB polarization map and its corresponding angular power spectrum for each case outlined in Table \ref{table_random}. }
	\label{fig:noiseless}
\end{figure*}

After training the network, we proceed to evaluate its performance on the test set. The outcomes of foreground removal by the network are presented in Figure \ref{fig:map_2sigma_lite}.  The residual maps demonstrate a successful removal of foreground contamination, exhibiting a clean separation between the foreground and CMB components. Moreover, the residual maps retain a greater amount of information in the galactic plane, which is a region heavily affected by foreground contamination. To assess the accuracy of the network, we calculate the average MAD values across 300 testing sets. The recovered Q map yields an average MAD of $0.029 \pm 0.004\ \mu$K, while the U map yields an average MAD of $0.032 \pm 0.009\ \mu$K. Additionally, we examine the recovery of the noisy EE and BB power spectra, as illustrated in the upper panels of Figure \ref{figure_s4_cross_QUEB_lite}. Obviously, the recovered noise EE and BB power spectra are very consistent with the target power spectrum, indicating that our neural network effectively removes foreground contamination. The EE and BB spectra experience a sharp increase at scales $\ell>600$ due to the impact of the instrumental beam and noise.

In order to mitigate the influence of noise on the power spectra, we employ the cross-correlation technique between two reconstructed HS maps. The results shown in the lower left panel of Figure \ref{figure_s4_cross_QUEB_lite} demonstrate a remarkable consistency between the EE angular power spectrum and the fiducial EE spectrum for $\ell\lesssim900$. To quantitatively assess the agreement, we calculate the coefficient $R^2$ for the denoised EE power spectrum. The obtained values of $R^2= 0.98\pm 0.02$ (68\% confidence level) for $\ell<900$  and $R^2= 0.999\pm 0.0005$ for $\ell<800$ indicate a strong concordance between the recovered EE power spectrum and the fiducial spectrum. However, it is important to note that the input maps utilized in the network model inevitably suffer from a loss of information regarding small-scale structures ($\ell>900$) due to the presence of an instrumental beam with a large FWHM of 28.9 arcmin. Consequently, the output map from the network also lacks this high-$\ell$ information. As a result, the network is unable to recover the EE spectrum for $\ell>900$.

Additionally, we present the recovery of the lensing B-mode power spectrum in the lower right panel of Figure \ref{figure_s4_cross_QUEB_lite}. Notably, we observe that the lensing BB power spectrum can be accurately reconstructed for angular scales $\ell<500$. To quantitatively evaluate the agreement, we calculate the coefficient $R^2$ for the denoised lensing BB power spectrum, yielding a value of $R^2= 0.95\pm 0.03$ (68\% confidence level) for $\ell<500$ and $R^2= 0.98\pm 0.01$ (68\% confidence level) for $\ell<400$. However, it is important to acknowledge that the lensing B-mode power spectrum for $\ell>500$ cannot be recovered due to the presence of instrumental noise and the effects of the instrumental beam, as discussed in Section \ref{CMB_S4_Experiment}.

\section{Discussion}
\subsection{Variation of the foreground parameters in the noiseless case}
\label{Noiseless_Case}
In Section \ref{simulations}, the simulation of foregrounds involves treating the amplitude $A$ and spectral index $\beta$ parameters for each foreground component as Gaussian random variables. However, given that CNN methods rely on simulated data in the training set, the randomization of parameters ($A$ and $\beta$) in terms of their size could impact the obtained results. Hence, it is necessary to investigate the influence of parameter variations on our findings. Table \ref{table_random} outlines five distinct cases considered for the variation of parameters ($A$ and $\beta$). For Case 1, no manual uncertainty is introduced to the amplitude $A$ and spectral index $\beta$. In Case 2, all pixel value in the amplitude template map $A$ (and spectral index map $\beta$) is multiplied by a random number drawn from a distribution with an average value of 1 and a standard deviation of 0.1 (0.05 for $\beta$). Similarly, in Case 3, the variation ranges are increased, and the multiplication factor for all pixel value in the amplitude template map $A$ (and spectral index map $\beta$) is a random number drawn from a distribution with an average value of 1 and a standard deviation of 0.15 (0.1 for $\beta$). It is important to note that for Cases 2 and 3, the same random number is multiplied for all pixel values in the $A$ (or $\beta$) map, implying that all pixel values in the $A$ (or $\beta$) map vary together. In Case 4, a map is generated where each pixel value independently follows a Gaussian distribution with a mean of 1 and a standard deviation of 0.1 for $A$ (0.05 for $\beta$), and the $A$ (or $\beta$) map is then multiplied by this generated random map. In Case 5, the variation ranges of $A$ and $\beta$ are increased compared to Case 4. The variation range in Case 4 and Case 5 is the same as in Case 2 and Case 3, respectively. However, for Cases 4 and 5, each pixel value in the $A$ and $\beta$ maps is multiplied by an independent random number, indicating that all pixel values in the $A$ and $\beta$ maps vary independently.
\begin{table*}
	\begin{center}
		\centering
		\caption{Cases of variation of the spectral parameters. $A$ and $\beta$ are amplitude and spectral index of foregrounds (see section \ref{data_foreground} ). }
		\label{table_random}
		\begin{tabular}{ c|c|c|c|c}
			\hline
			& $A$ & $\beta$ & Random pixel independently & MAD of the recovered Q map [$\mu K$]  \\
			\hline
			Case 1 & 0\% & 0\% & - & $0.0025\pm 0.0001$ \\
			\hline
			Case 2 & 10\% & 5\% & No & $0.0061 \pm 0.0004$  \\
			\hline
			Case 3 & 15\% & 10\% & No & $0.0091 \pm 0.0012$ \\
			\hline
			Case 4 & 10\% & 5\% & Yes & $0.0112 \pm 0.0012$ \\
			\hline
			Case 5 & 15\% & 10\% & Yes & $0.0174 \pm 0.0014$\\
			\hline
		\end{tabular}
	\end{center}
\end{table*}

\begin{figure*}
	\centering
	\includegraphics[width=0.7\hsize]{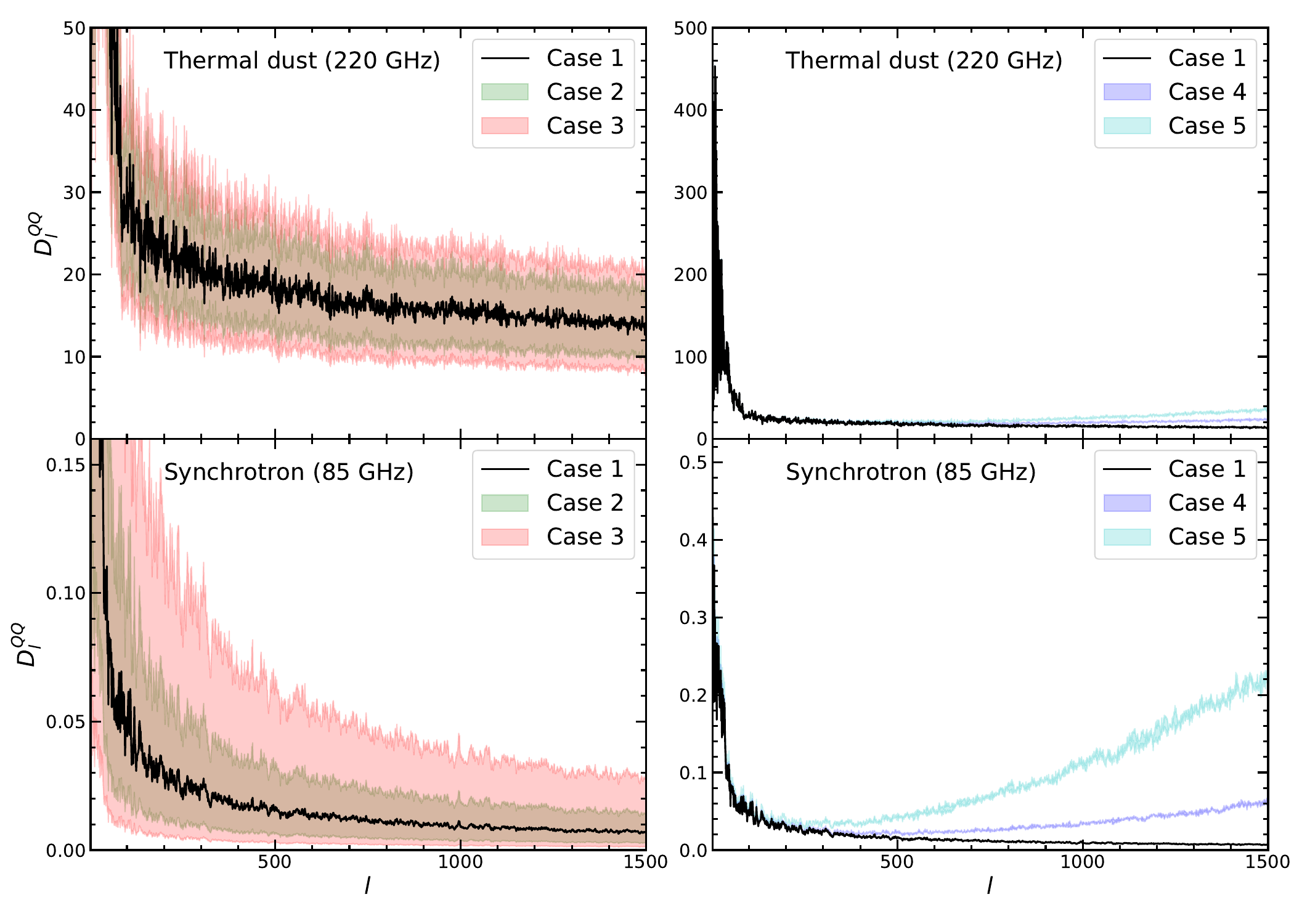}
	\caption{The angular power spectra of simulated thermal dust and synchrotron maps for cases 1, 2, 3, 4, and 5 outlined in Table \ref{table_random}. The black lines refer to the power spectra of the template. For the cases 2-5, corresponding shaded areas show the $\pm 2\sigma$ ranges of simulated power spectra determined from 200 different random realization.}  
	\label{figure_power_syn}
\end{figure*}

In this section, we assess the impact of foreground parameter variations on the removal of foreground components. We conduct our evaluation by applying our method to simulated data that emulates the performance of the CMB-S4 experiment. Table \ref{table_random} presents five distinct random cases, characterized by different standard deviation sizes and forms of random realizations. For each random case, we generate 1000 beam-convolved emission maps at eight frequency bands, with a resolution parameterized by ${\rm N_{side}}$ of 512. The frequency bands and instrumental properties of the CMB-S4 experiment are summarized in Table \ref{table_1}. To accurately investigate the effect of changes in foreground parameters on the network model, we exclude noise from the simulated data. The test set comprises 300 sky emission sets, similar to the training sets, but with different random seeds and parameter values. The inputs to the network model consist of the beam-convolved observational maps ($Q$ maps or $U$ maps) at the eight frequencies, encompassing the CMB and foregrounds. The desired output of the network is a beam-convolved CMB map ($Q$ map or $U$ map) with a beam at 220 GHz. Each random case is trained using a separate network.

The outcomes for each random case are shown in Figure \ref{fig:noiseless}. Notably, for case 1, the residual map exhibits a high degree of cleanliness. The MAD between the recovered CMB Q map and the true Q map is measured to be $0.0025\pm 0.0001\ \mu$K. Furthermore, the angular power spectrum of the Q map can be accurately recovered, indicating effective removal of foregrounds.  For random cases 2-5, the residual maps appear to retain more information compared to case 1. The MADs for these cases are determined to be $0.0061\pm 0.0004\ \mu$K, $0.0091\pm0.0012\ \mu$K, $0.0112\pm 0.0012\ \mu$K, and $0.0174\pm 0.0014\ \mu$K, respectively. Analysis of the recovered CMB maps reveals that changes in foreground parameters can degrade the performance of the network. Particularly, cases involving random pixel independence (cases 4 and 5) demonstrate a more significant degradation in performance of the network compared to cases involving random pixel dependence (cases 2 and 3).

Subsequently, we compute the power spectrum of the thermal dust and synchrotron Q maps for each random case. As depicted in Figure \ref{figure_power_syn}, the simulated power spectra of the thermal dust and synchrotron for random cases 2 and 3 adequately align with the template power spectra (case 1) across all considered scales. However, for random cases 4 and 5, the simulated power spectra deviate from the template power spectra, particularly at small scales. These deviations resemble noise, potentially suggesting that the random realizations in cases 4 and 5 could introduce noise \citep{Wang:2022}. Consequently, we think that the noise stemming from random cases 4 and 5 will significantly impact the recovery of the CMB map. 

Finally, it is important to acknowledge that the discussion in this section does not account for the presence of instrumental noise. A comparison with the results showed in Figure \ref{fig:map_2sigma}  reveals that the residual map is notably cleaner when noise is not taken into consideration. This observation highlights the detrimental effect of noise on the accuracy and quality of our recovered map results. 

\begin{figure*}
	\centering
	\subfigure[Case 2]{
		\includegraphics[width=17cm]{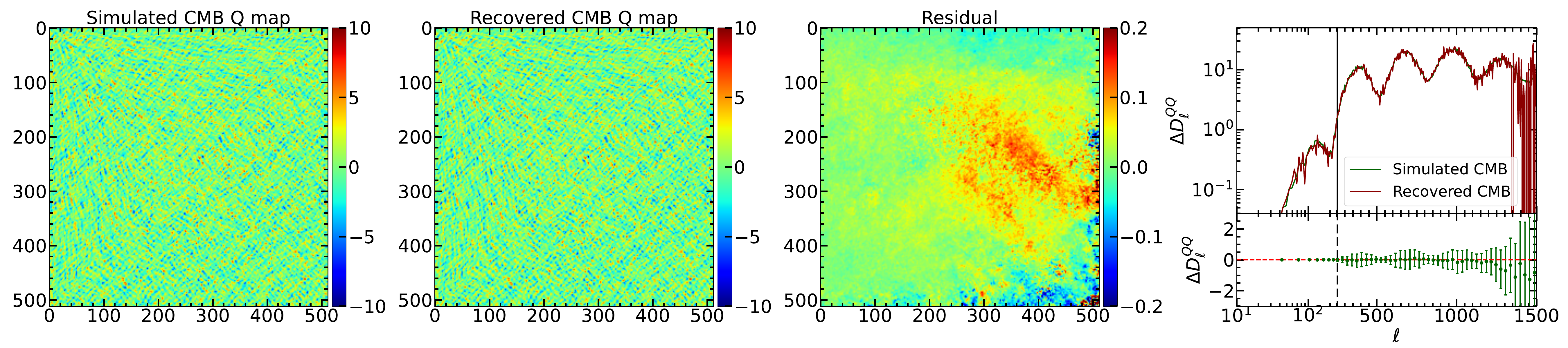}
	}
	\vspace{-1mm}
	\subfigure[Case 3]{
		\includegraphics[width=17cm]{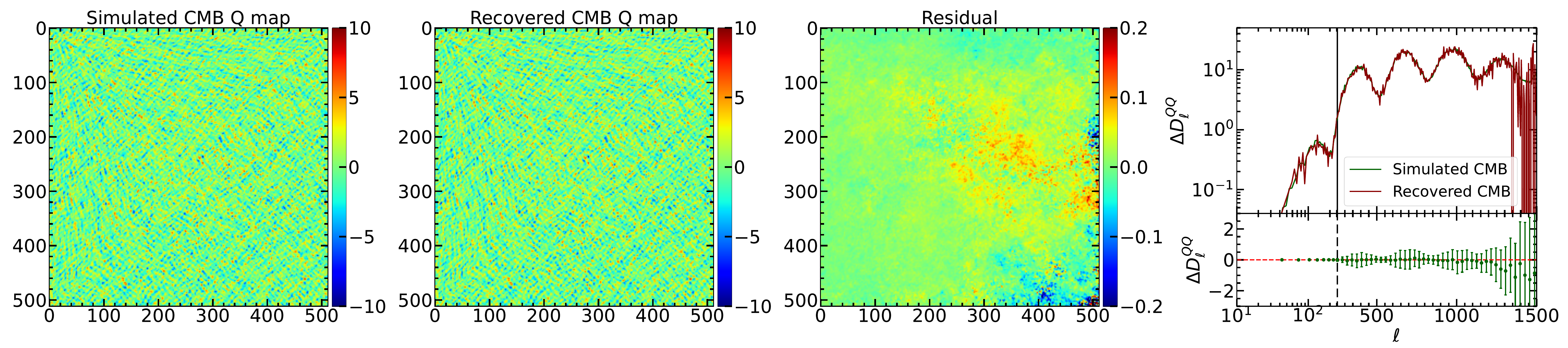}
	}
	\vspace{-1mm}
	\subfigure[Case 4]{
		\includegraphics[width=17cm]{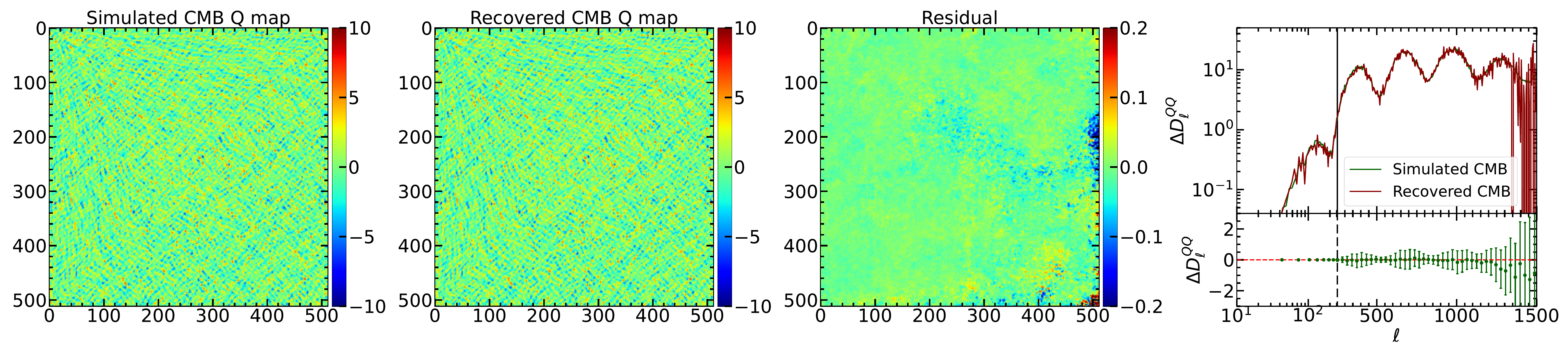}
	}
	\subfigure[Case 5]{
		\includegraphics[width=17cm]{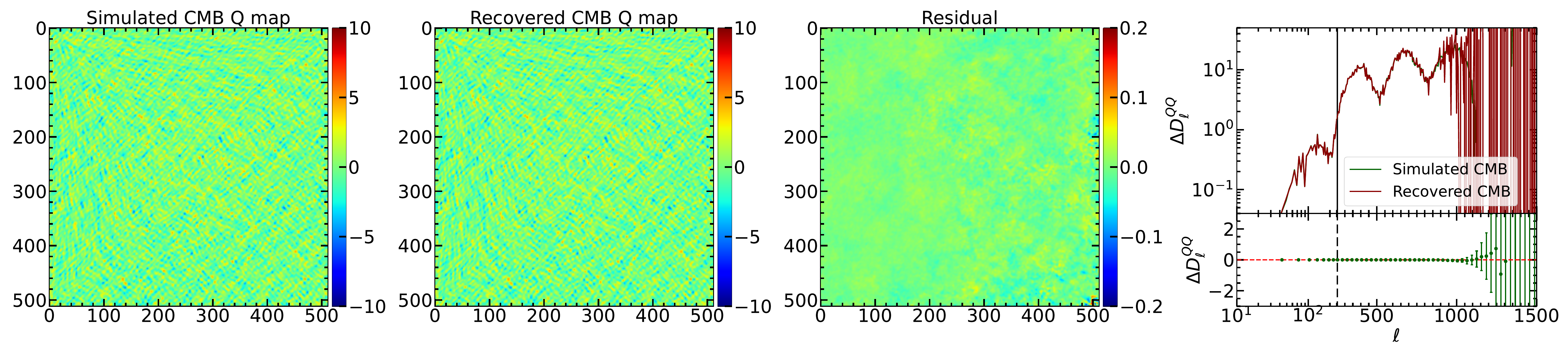}
	}
	\caption{Recovered noisy CMB map and its corresponding angular power spectrum for varying the number of frequency bands in the input data.  The recovered QQ power spectra presented in the analysis are obtained after the denoising procedure. Each case is listed in Table \ref{table_freq}. }
	\label{fig:multi_freq}
\end{figure*}

\subsection{Frequency selection}
\label{freq_sele}
In Section \ref{CMB_S4_Experiment}, we employ our proposed methodology to analyze simulated data corresponding to the performance of the CMB-S4 experiment. Specifically, we utilize beam-convolved observational maps (either $Q$ maps or $U$ maps) obtained from eight different frequency bands as inputs to the network model. The desired output of the network is a beam-convolved CMB map ($Q$ map or $U$ map) with both noise and beam effects at 220 GHz. In this section, we investigate the impact of varying the number of frequency bands in the input data on the obtained results, thereby elucidating the influence of frequency selection.

\begin{table*}
	\begin{center}
		\centering
		\caption{The impact of varying the number of frequency bands in the input data on the performance of the network model.}\label{table_freq}
		\begin{tabular}{ c|c|c|c}
			\hline
			&Input frequencies & Output frequency & MAD of the recovered Q map    \\
			&(GHz) &(GHz) &[$\mu K$]  \\
			\hline
			Case 1&30,40,85,95,145,155,220,270 & 220  & 0.016$\pm$ 0.008 \\
			\hline
			Case 2&220 & 220 & $0.03\pm 0.013$  \\
			\hline
			Case 3&155,220 & 220 & $0.027\pm 0.010$  \\
			\hline
			Case 4&85,95,155,220 & 220 & $0.023\pm 0.007$  \\
			\hline
			Case 5&30,40,85,95,145,155,220,270 & 155  & $0.012\pm 0.003$ \\
			\hline
		\end{tabular}
	\end{center}
\end{table*}

In order to facilitate comparison, we consider five distinct cases as enumerated in Table \ref{table_freq}. In Case 1, the network model takes as input the beam-convolved observational maps ($Q$ maps or $U$ maps) from eight different frequency bands, while the desired output of the network is a beam-convolved lensed CMB map ($Q$ map or $U$ map) with noise and beam effects at 220 GHz. It should be noted that Case 1 aligns with the methodology employed in Section \ref{CMB_S4_Experiment}. Moving on to Case 2, we only utilize the observational maps at 220 GHz as inputs to the network model, while the desired output remains unchanged. In Case 3, the input data consists of the observational maps at 155 GHz and 220 GHz, while the desired output remains consistent. Similarly, in Case 4, the network model takes as input the observational maps at 85 GHz, 95 GHz, 155 GHz, and 220 GHz, while the desired output remains unaltered. In Case 5, the input data comprises the observational maps from all eight frequencies, while the desired output is a beam-convolved lensed CMB map ($Q$ map or $U$ map) with noise and beam effects taken into account at 155 GHz. Notably, the desired output for Case 5 is convolved with a larger Gaussian beam of FWHMs = 22.7 arcmin compared to the beam of FWHMs = 13.0 arcmin at 220 GHz, thereby resulting in a higher degree of smoothing and loss of fine-scale information on the map. The selection of the Case 5 is made to assess the influence of an increased beam size on the network output. It is worth mentioning that the training and test sets for all cases are derived from Section \ref{CMB_S4_Experiment}, and each case is trained using a separate network.

Figure \ref{fig:multi_freq} illustrates the outcomes of the noisy CMB Q map recovery on the test set for Cases 2-5, while the recovery results for Case 1 are depicted in Figure \ref{fig:map_2sigma}. It is important to note that, for the sake of brevity, we only present the results of Q map recovery, although the U map recovery yields similar outcomes.  We can observe that the residual maps progressively exhibit cleaner features from Case 2 to Case 4, and ultimately to Case 1. This suggests that the recovery of the CMB noisy Q map improves as the number of frequency bands in the input data of the network increases. This can be attributed to the fact that the multi-band data provides a greater wealth of foregrounds and CMB signal information to the network, thereby facilitating more effective foreground removal. To provide a quantitative assessment, we calculate the average MAD values across 300 testing sets for each case listed in Table \ref{table_freq}: $0.016 \pm 0.008\ \mu$K for Case 1, $0.03 \pm 0.013\ \mu$K for Case 2, $0.027 \pm 0.010\ \mu$K for Case 3, and $0.023 \pm 0.007\ \mu$K for Case 4. The comparison between these MAD values further supports our conclusion. Furthermore, Figure \ref{fig:multi_freq} depicts the recovered CMB QQ after the denoising step. We can see that QQ power spectra can be accurately recovered, indicating that the number of frequency bands in the input data of the network has negligible influence on the power spectrum recovery.

The average MAD values for Case 5 are computed as $0.012 \pm 0.004\pm 0.004 \mu$K, slightly smaller than the MAD values observed in Case 1. This suggests that, in terms of map-level recovery, the efficacy of Case 5 surpasses that of Case 1. This phenomenon could be due to the discrepancy in sensitivity between the 155GHz and 220GHz frequency bands. Specifically, the output of the Case 5 exhibits lower noise levels compared to the output of the Case 1 due to a lower sensitivity of the 155GHz frequency band. 
However, due to the larger beam size in the target map, the recovered map also suffers from a loss of small-scale information, which is evident in the recovered power spectrum. For Case 5, Figure \ref{fig:multi_freq} demonstrates that the QQ power spectra can be accurately recovered for $\ell<1100$. However, it is important to note that the uncertainty in the recovered QQ spectrum significantly increases as a consequence of the larger beam effect present in the training target map.

\begin{figure*}
	\centering
	\subfigure[Q map]{
		\includegraphics[width=17cm]{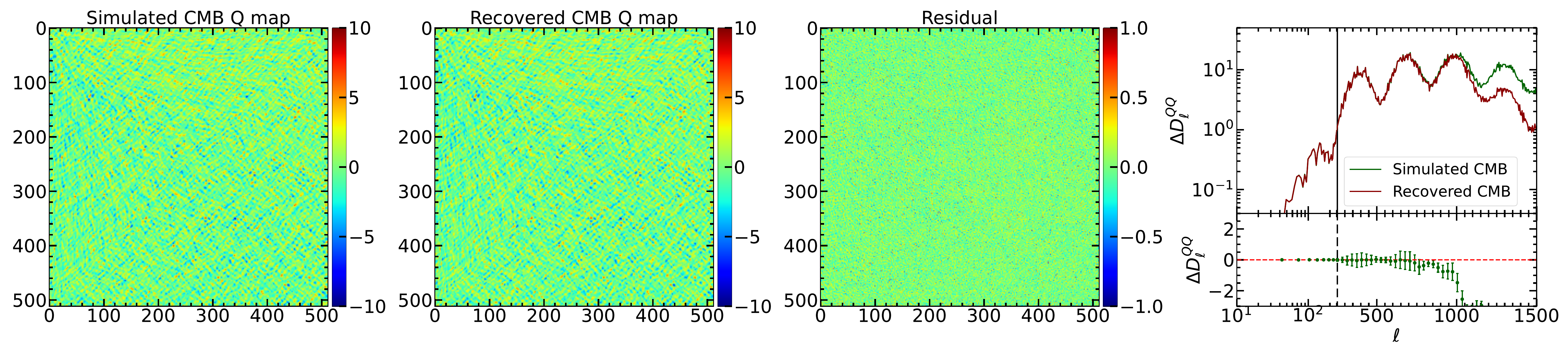}
	}
	\vspace{0mm}
	\subfigure[U map]{
		\includegraphics[width=17cm]{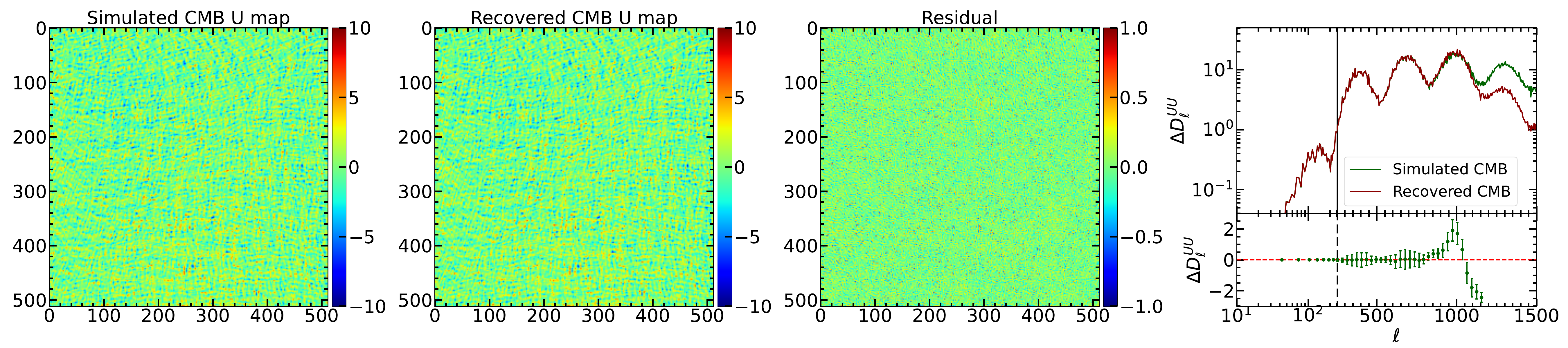}
	}
	\caption{Recovered CMB polarization Q maps and their angular power spectra for the partial-sky observation data with the performance of CMB-S4 experiment. Here, the network model is designed to remove the foregrounds and noise. The simulated maps utilized in this analysis comprise the beam-convolved CMB map without noise map. The recovered maps correspond to the pure CMB maps reconstructed using the {\tt CMBFSCNN} algorithm.}
	\label{fig:s4_denoise_Q}
\end{figure*}

\begin{figure*}
	\centering
	\includegraphics[width=1\hsize]{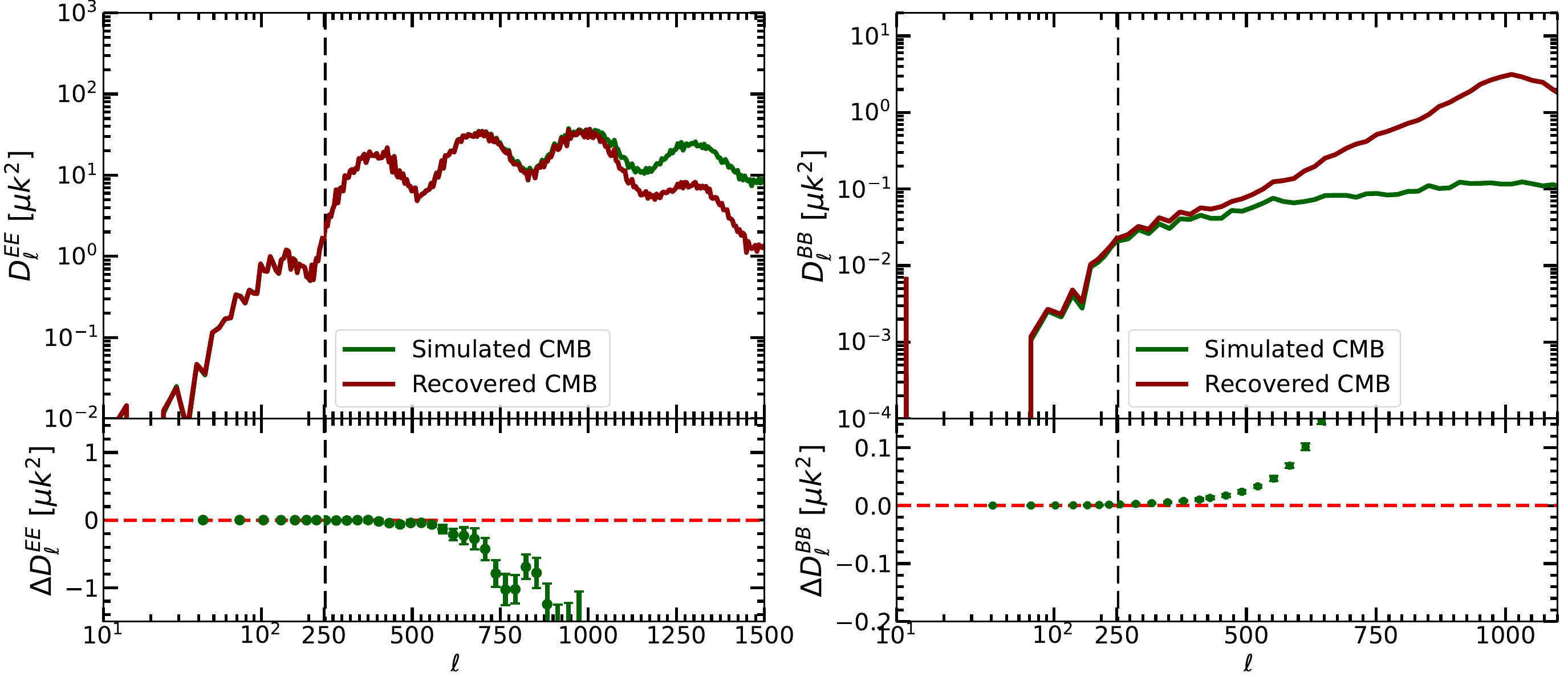}
	\caption{The recovery of EE and lensing B-model power spectra from the recovered pure CMB Q/U maps as depicted in Figures \ref{fig:s4_denoise_Q}.}  
	\label{figure_s4_denoise_cross_QUEB}
\end{figure*}
Considering the findings from the analysis in Section \ref{Noiseless_Case}, it can be concluded that the presence of instrumental noise and beam effects significantly impairs the accuracy of the recovered results, particularly at smaller scales. Consequently, for the purpose of recovering the CMB signal, the 220GHz beam and noise are selected as the instrument characteristics of target map in Section \ref{CMB_S4_Experiment}. This choice is driven by multiple factors, including its smaller FWHM  in comparison to other lower frequency bands, as well as the lower noise levels in comparison to the 270GHz frequency. These considerations are anticipated to enhance the network's ability to recover the CMB signal. Similarly, for the LiteBIRD experiment in Section \ref{lite_exp}, the beam at the 166GHz frequency band is chosen as the beam of target map, considering both noise and beam characteristics, as it yields the most favorable outcome.

\subsection{Map Denoising}
\label{map_denoise}

In Section \ref{CMB_S4_Experiment}, we have demonstrated the efficacy of our network in effectively eliminating foregrounds. However, it is important to note that the output CMB maps generated by the neural network still retain the presence of instrumental noise. To address this concern, we employed a cross-correlation technique to mitigate the impact of instrumental noise on the power spectra. In this section, our objective is to employ the network model to remove the noise at the level of the CMB map, using simulated data with the performance of the CMB-S4 experiment. Consequently, we configured the inputs to the network as the beam-convolved observational maps at the eight frequencies, including the CMB signal, foregrounds, and instrumental noise. Meanwhile, the  outputs of the network is the beam-convolved CMB map without instrumental noise, meaning that our network has been designed to  remove both foregrounds and instrumental noise components.

Figure \ref{fig:s4_denoise_Q} show the outcomes obtained from the test set using simulation data for the CMB-S4 experiment. Notably, the Q and U residual maps retain a significant amount of information, suggesting that the accurate recovery of the pure CMB Q/U maps remains challenging. The recovered $QQ$ and $UU$ power spectra align closely with the simulated counterparts for $\ell \lesssim 900$, but deviate gradually as the multipole moments increase beyond $\ell > 900$. Additionally, we compute the EE and BB power spectra from the recovered CMB Q/U maps, as illustrated in Figure \ref{figure_s4_denoise_cross_QUEB}. The EE power spectrum demonstrates consistency with the fiducial spectrum for $\ell \lesssim 900$, but exhibits increasing deviation for higher multipoles at $\ell > 900$. Moreover, the lensing B-model power spectrum exhibits a gradual deviation as the multipoles increase beyond $\ell > 300$. These findings suggest that our network model struggles to accurately recover information at small scales. Consequently, distinguishing between the polarized CMB and noise at the map level proves to be a challenge for the network model.

\subsection{Dependency of foreground models}
\label{sec:depend_fg}

\begin{figure*}
	\centering
	\includegraphics[width=1.0\hsize]{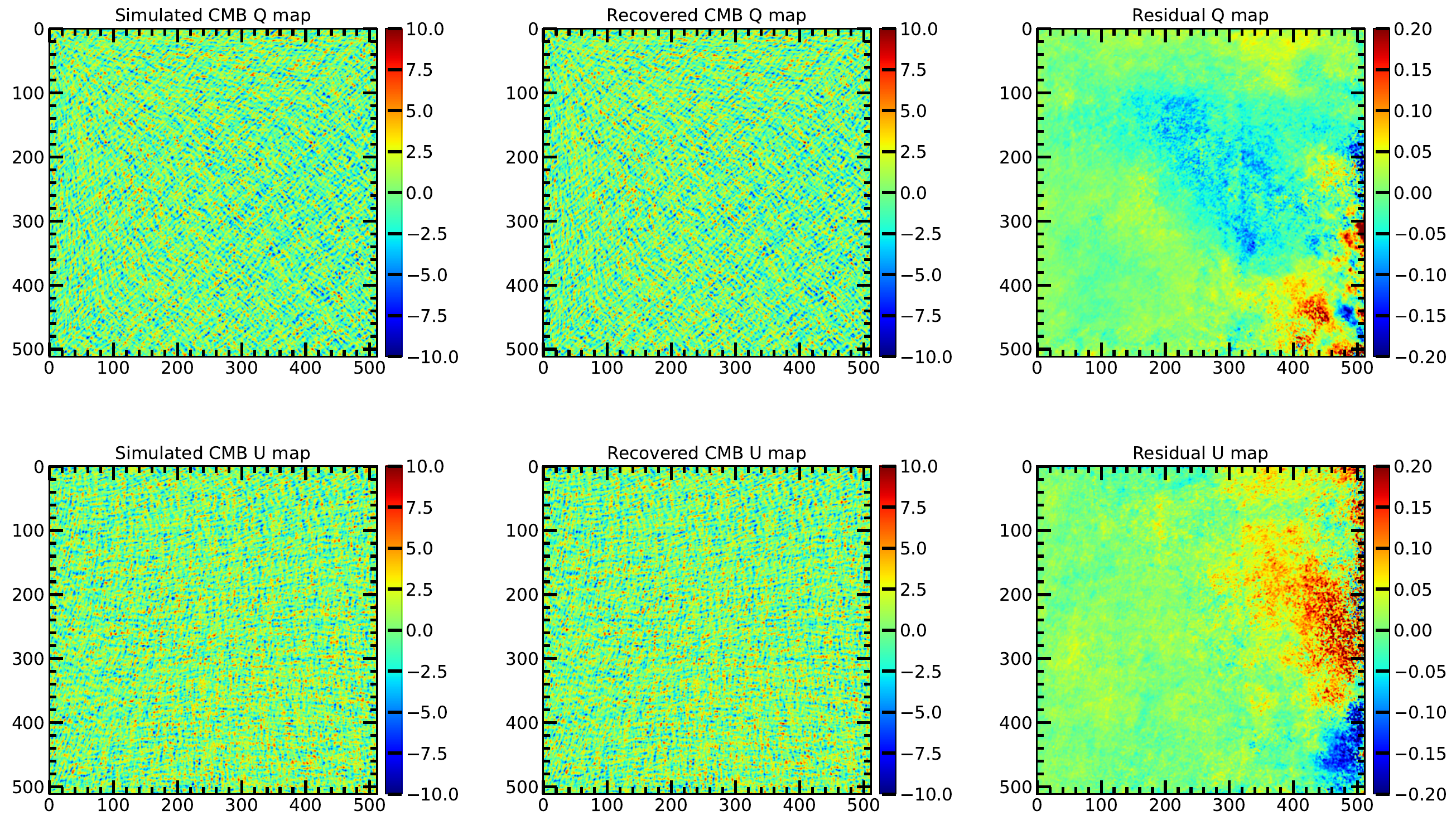}
	\caption{The recovered noisy CMB polarization maps for the \texttt{Experiment 1}.  The simulation of the synchrotron radiation in the test set is based on the s2 model. }  
	\label{fig:map_2sigma_s2}
\end{figure*}
\begin{figure*}
	\centering
	\includegraphics[width=1\hsize]{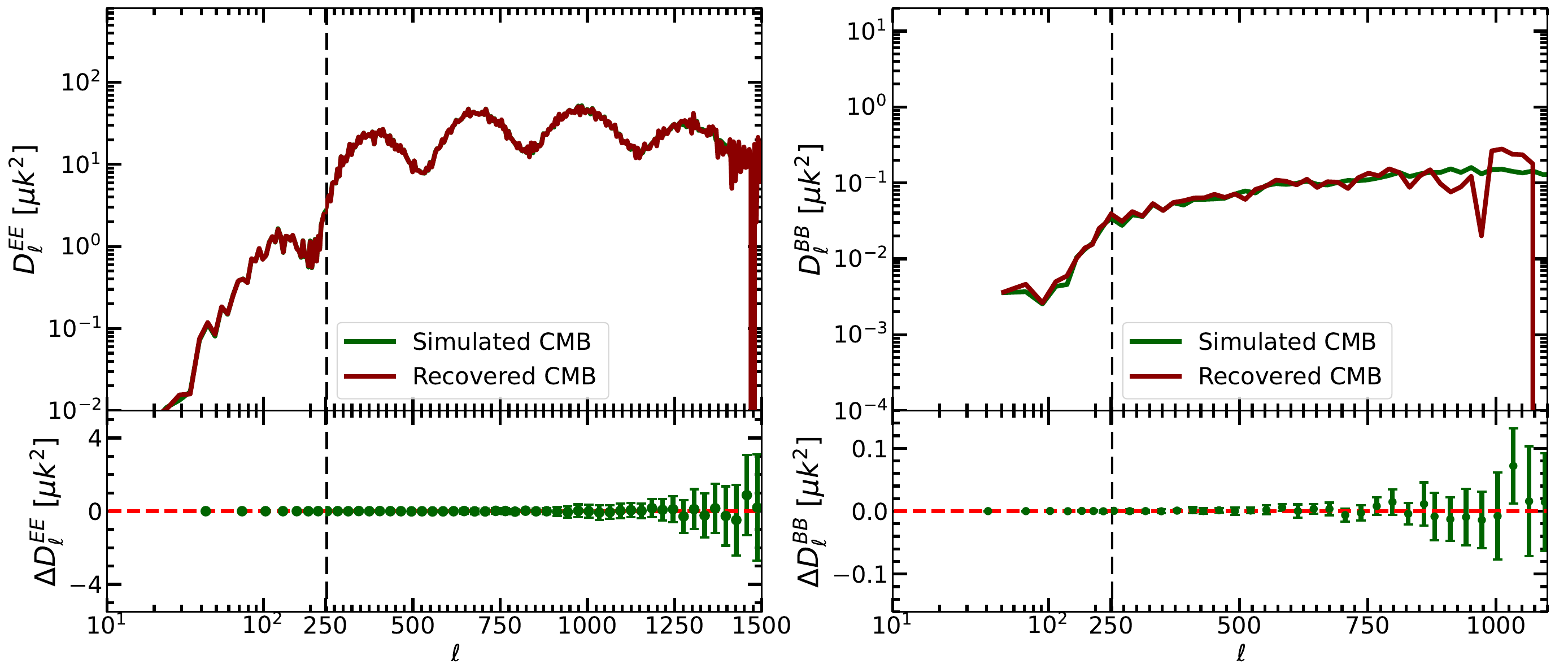}
	\caption{The recovery of EE and lensing B-model power spectra from the recovered CMB Q/U maps as depicted in Figures \ref{fig:map_2sigma_s2}.}  
	\label{figure_s4_denoise_cross_QUEB_s2}
\end{figure*}

\begin{figure*}
	\centering
	\includegraphics[width=1.0\hsize]{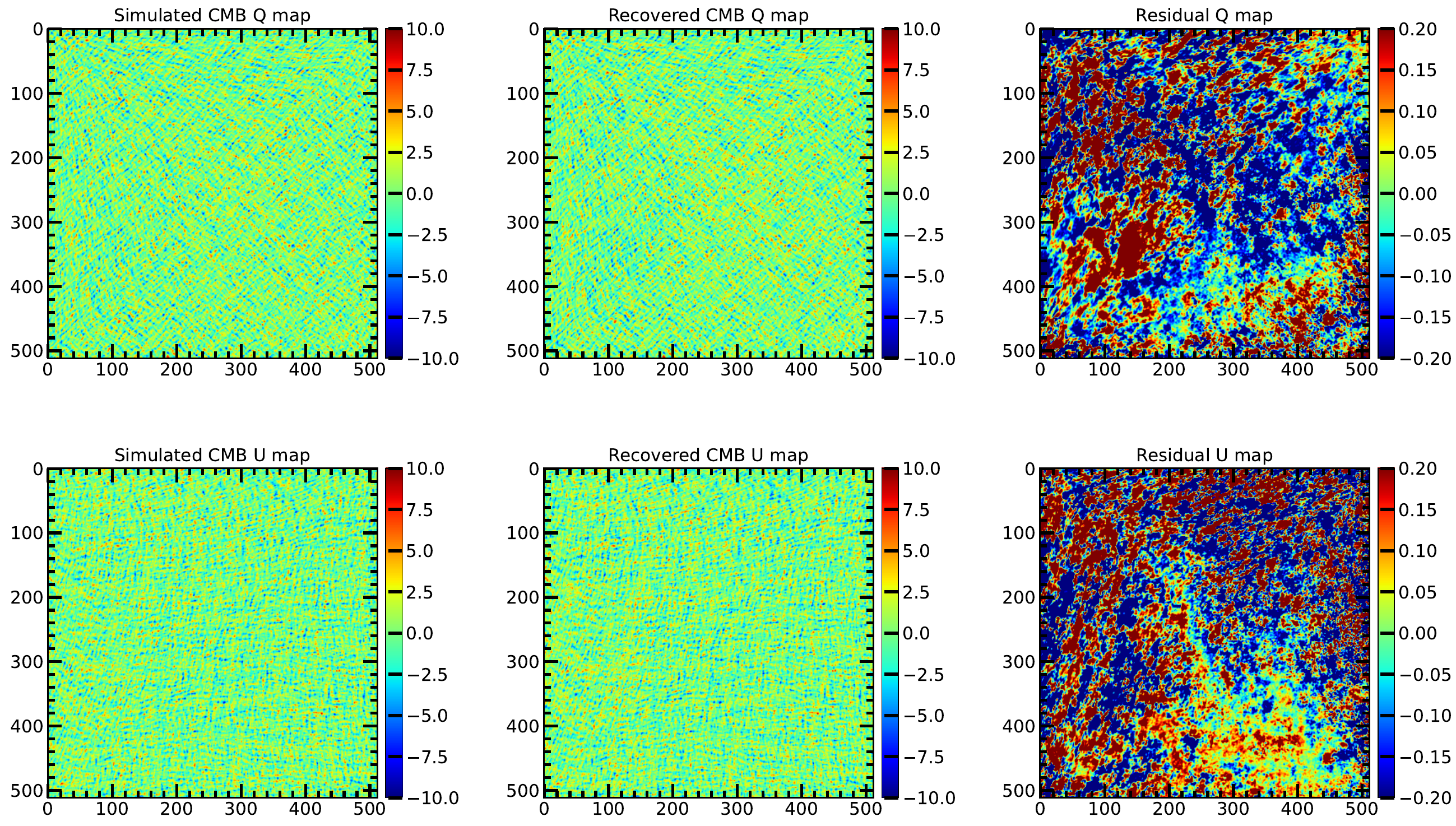}
	\caption{The recovered noisy CMB polarization maps for the \texttt{Experiment 2}.  The simulation of the synchrotron radiation and thermal dust radiation in the test set is based on the s2 and d4 model. }  
	\label{fig:map_2sigma_s2d2}
\end{figure*}
\begin{figure*}
	\centering
	\includegraphics[width=1\hsize]{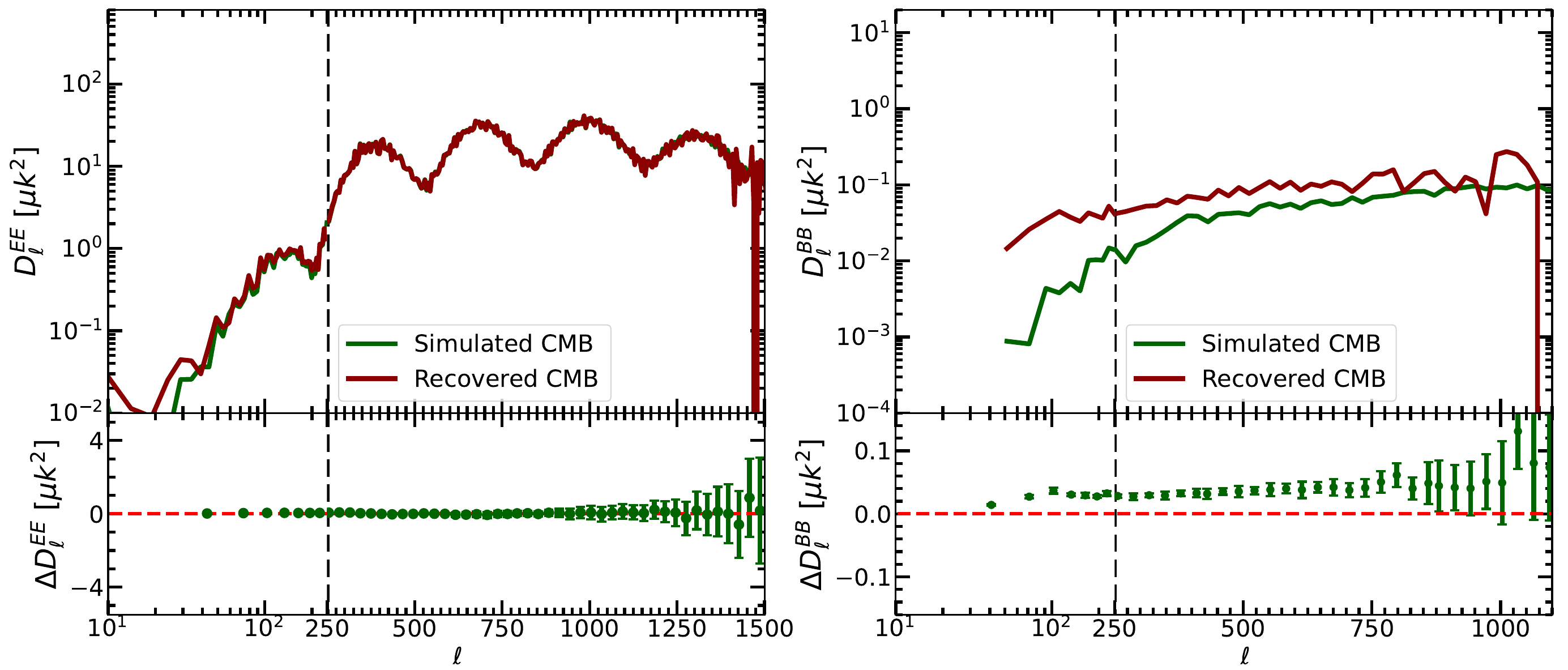}
	\caption{The recovery of EE and lensing B-model power spectra from the recovered CMB Q/U maps as depicted in Figures \ref{fig:map_2sigma_s2d2}.}  
	\label{figure_s4_denoise_cross_QUEB_s2d2}
\end{figure*}
\begin{figure*}
	\centering
	\includegraphics[width=1\hsize]{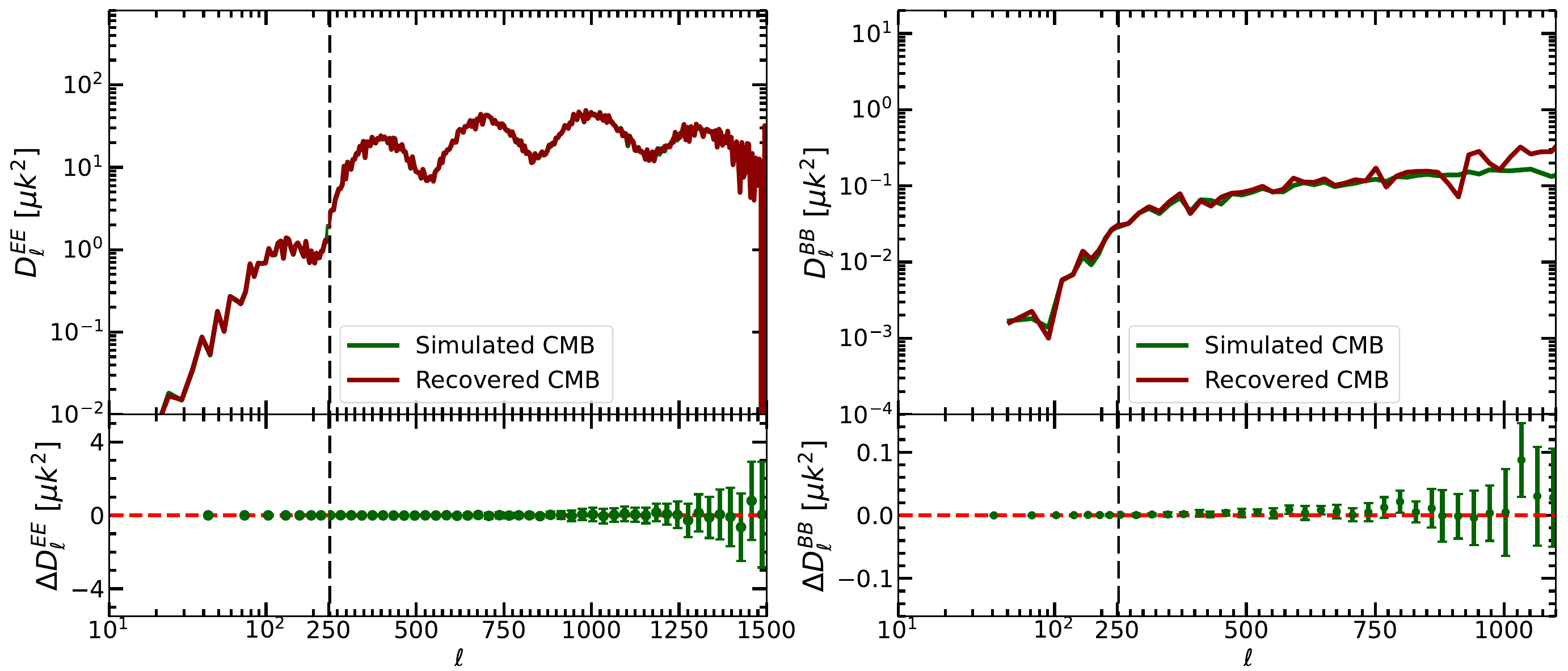}
	\caption{The recovery of EE and lensing B-model power spectra from test set, where the simulation of the synchrotron radiation and thermal dust radiation is based on the s2 and d4 model. Here, the CNN model is trained in the training set, which includes sky maps of the synchrotron radiation s2 model and the dust radiation d4 model.}  
	\label{figure_s4_denoise_cross_QUEB_s2d2_tra}
\end{figure*}

Our CNN model training relies on the training set, where the simulation of the foregrounds is based on the models. Therefore, our method is a parameterized component separation algorithm. Our CNN model will inevitably depend on the foreground model. Here, we test the dependency of the CNN model on foreground models. We train the network on a original training set from Section \ref{CMB_S4_Experiment}. Then, we vary the foreground models in the test set to examine the performance of the network model under different foreground models. Specifically, the simulation of synchrotron radiation and thermal dust radiation in the training set is based on the s1 and d1 models from the \texttt{PySM} package. However, the simulation in the test set is based on the synchrotron radiation s2 or thermal dust radiation d4 models. Compared to the synchrotron radiation s1 model, the s2 model takes into account the spectral index varying with latitude. The spectral index of s2 model is defined as $\beta_\mathrm{s2}=\beta_\mathrm{s,b=0}+\delta_\beta\sin|b|$, where $b$ are the Galactic latitude, respectively. We use a gradient $\delta_\beta=-0.3$ based on the WMAP polarization data. Compared to the the thermal dust d1 model, d4 model has the two dust components.

Here we present two experiments. The first experiment involves adjusting the synchronous radiation model, denoted as \texttt{Experiment 1}. The training dataset is derived from the simulations detailed in Section \ref{CMB_S4_Experiment}, where data was simulated to reflect the capabilities of the CMB-S4 experiment. In generating test set of \texttt{Experiment 1}, synchronous radiation model was substituted with the s2 model. In the second experiment, denoted as \texttt{Experiment 2}, both the synchronous radiation model and the thermal dust model were simultaneously altered. When creating test set of \texttt{Experiment 2}, synchronous radiation and the thermal dust model were replaced by the s2 model and d4 model, respectively. Once the network model is well-trained on the training set, we input the test sets of \texttt{Experiment 1} and \texttt{2} into the trained network model. 

We firstly present results of \texttt{Experiment 1}. As shown in the Figure \ref{fig:map_2sigma_s2}, a small amount of information remains in the residual maps, indicating that the foregrounds can be cleanly removed. The recovered Q map yields an average MAD of $0.032 \pm 0.016\ \mu$K, while the U map yields an average MAD of $0.030 \pm 0.024\ \mu$K. These MAD values are slightly larger than the MAD values from the Section \ref{CMB_S4_Experiment}, where the MAD values in the Section \ref{CMB_S4_Experiment} are $0.016 \pm 0.008\ \mu$K for the Q map recovery and $0.021 \pm 0.002\ \mu$K for the U map recovery. Figure \ref{figure_s4_denoise_cross_QUEB_s2} displays the CMB power spectrum we reconstructed on the test set of \texttt{Experiment 1}, We can observe that both the CMB EE power spectrum and the lensing BB power spectrum can be accurately recovered, consistent with the results from Section \ref{CMB_S4_Experiment}. These results indicate that altering the synchronous radiation model has a minimal impact on our results.

Then, we present results of \texttt{Experiment 2}.  As shown in the Figure \ref{fig:map_2sigma_s2d2}, the residual map retains a amount of information, indicating that there exist significant foreground residuals in the reconstructed noisy CMB Q/U map. We calculate the average MAD values across testing set. The recovered CMB Q map yields an average MAD of $0.195 \pm 0.049\ \mu$K, while the U map yields an average MAD of $0.241 \pm 0.059\ \mu$K. These MAD values are about ten times greater than the values in Section \ref{CMB_S4_Experiment}.  Figure \ref{figure_s4_denoise_cross_QUEB_s2d2} displays the CMB power spectrum we reconstructed on the test set of \texttt{Experiment 2}. It can be observed that residual foreground effects exhibit a slight impact on the EE power spectrum at angular scales $\ell<200$, but their influence at large scales can be neglected. Therefore, our CNN model remains effective in recovering the CMB EE power spectrum. However, residual foreground effects significantly influence the recovery of the lensing BB power spectrum, leading to noticeable bias in the recovered BB power spectrum across all angular scales.

Based on the results of \texttt{Experiment 1} and \texttt{Experiment 2}, it is evident that altering the model of thermal dust radiation has a significant impact on the recovery of lensing B-mode. This also demonstrates the dependency of our method on foreground models. Although we randomized the parameters of the foreground models during the simulation of the training set, the lack of inclusion of different foreground models in the training set resulted in a higher amount of foreground residual in reconstructed CMB maps when changing the thermal dust radiation model. Given the strong fitting capability of CNN methods, we can incorporate various thermal dust and synchronous radiation models into the training set. That is, during the simulation of the training set, generalize the foreground models as well. This approach can effectively alleviate the issue of CNN methods relying on foreground models.

We augmented the original training set with 300 sets of multi-frequency observed sky maps. The simulations of synchrotron and thermal dust emissions in these 300 sky maps were based on the s2 and d4 models. During the simulation process, we still randomized the spectral indices and amplitudes of foregrounds. We trained the network model on this expanded training set. After completing the network training, we input the test set into the trained network. The simulations of synchrotron and thermal dust emissions are based on the s2 and d4 model in test set. The recovered CMB Q map yields an average MAD of $0.031 \pm 0.014\ \mu$K, while the CMB U map yields an average MAD of $0.038 \pm 0.029\ \mu$K. These MAD values are slightly larger than the MAD values from the Section \ref{CMB_S4_Experiment}. The results of recovering the CMB polarization power spectra are shown in the Figure \ref{figure_s4_denoise_cross_QUEB_s2d2_tra}, demonstrating that we can accurately recover the CMB EE and lensing BB power spectra. These results indicate that if the training set includes data from more foreground models, the network model trained on the training set can handle more complex foreground contamination.

\begin{figure*}
	\centering
	\includegraphics[width=1\hsize]{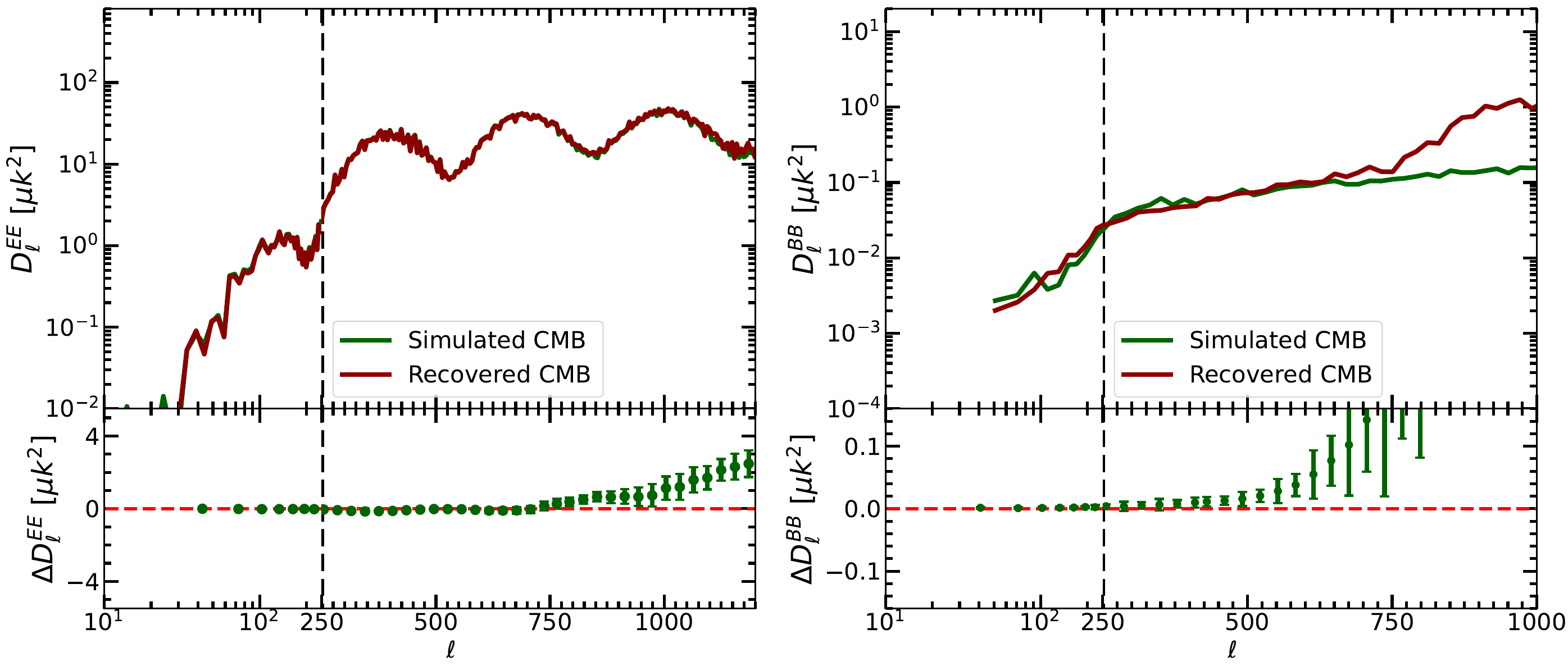}
	\caption{The recovery of EE and lensing B-model power spectra using NILC method.}  
	\label{figure_s4_denoise_NILC}
\end{figure*}

\subsection{Needlet domain ILC}
\label{sec:NILC}

ILC is a blind source method that has been widely applied to CMB foreground subtraction. As shown in equation (\ref{eq:ilc_QU}), the clean CMB map is represented as a weighted linear combination of sky maps observed in multiple frequency bands, with the weights calculated through a variance-minimization method.  Here, we employ ILC method to analyze simulated data corresponding to the performance of the CMB-S4 experiment. We compare the efficiency of our CNN method with the ILC method in removing polarized foregrounds.

Here, we use needlet domain ILC (NILC) method to recover CMB polarized maps.  The NILC method is an improvement of the ILC method and has been utilized for foreground removal in CMB temperature (T), polarization E-mode, and polarization B-mode maps\citep{Basak:2012,Basak:2013,Planck:2020,Zhang:2022,Dou:2023,Zhang:2024}. Firstly, we provide a brief overview of the NILC method.

We consider multi-frequency observational sky maps ($X^{\mathrm{\rm obs},\nu}(p)$) with varying instrument beams for each frequency band. The index $\nu$ and $p$ denote frequency and pixel, respectively. The observed maps convolves/deconvolutes to the same resolution in harmonic space: 
\begin{align}
\label{eq:conv}
X_{\ell m}^{\nu}=\frac{b_{\ell}}{b_{\ell}^{\nu}} X_{\ell m}^{\mathrm{\rm obs},{\nu}},
\end{align}
$X_{\ell m}^{\mathrm{\rm obs},{\nu}}$ is harmonic coefficients of maps ($X^{\mathrm{\rm obs},\nu}$). $b_{\ell}^{\nu}$ and $b_{\ell}$ represent beam window function for each frequency band and the common beam window function, respectively. After the correction of beam window function, we can assume that the CMB is frequency-independent. The maps are given as:
\begin{align}
	X^{\nu}(p)=X^{{\rm CMB}}(p) + X^{\nu,{\rm Fg}}(p)+ n^{\nu}(p),
\end{align}
$X^{{\rm CMB}}(p)$ and $X^{\nu,{\rm Fg}}(p)$ represent CMB map and foreground map, respectively. $n^{\nu}(p)$ is instrumental noise.

Each of these maps $X_{\ell m}^{\nu}$ with the common beam can be decomposed into a set of filtered maps $X_{\ell m}^{\nu,j}$ using filters $h_{\ell}^{j}$,
\begin{align}
X_{\ell m}^{\nu,j}=h_{\ell}^jX_{\ell m}^{\nu}.
\end{align}
The filters are chosen in such a way that
\begin{align}
\sum_j\left(h_l^j\right)^2=1.
\end{align}
In this work, we adopt the filters in the following form:
\begin{align}
\left.h_l^j=\left\{\begin{array}{cc}\cos\left[\left(\frac{l_{\rm mid}^j-l}{l_{\rm mid}^j-l_{min}^j}\right)\frac{\pi}{2}\right]&\text{for }l_{min}^j\leqslant l<l_{\rm mid}^j,\\\\\cos\left[\left(\frac{l-l_{\rm mid}^j}{l_{max}^j-l_{\rm mid}^j}\right)\frac{\pi}{2}\right]&\text{for }l_{\rm mid}^j<l\leqslant l_{max}^j\end{array}\right.\right.
\end{align}
In terms of $h_l^j$, the spherical needlets in HEALPix pixelization space are defined as
\begin{align}
\psi_{jk}(p)=\sqrt{\frac{4\pi}{N_j}}\sum_{\ell m}h_\ell^jY_{\ell m}(p)Y_{\ell m}^*(n_{jk}),
\end{align}
here, $N_j$ and $n_{jk}$ represent the number of pixels and $k-$th pixel of $j-$th needlet map, respectively. The needlet transformed coefficients for observed maps ($X^{\nu}$) are denoted as
\begin{align}
	\beta_{jk}^{\nu}&=\quad\int_{S^2}X^{\nu}(\hat{n}) \Psi_{jk}(\hat{n}) d\Omega_{\hat{n}}\\&=\quad\sqrt{\frac{4\pi}{N_j}}\sum_{l=0}^{l_{\max}}\sum_{m=-l}^{l}h_{l}^{j} X^{\nu}_{lm} Y_{lm}(\xi_{jk}),
\end{align}
its inverse transformation is given by
\begin{align}
\label{eq:inver}
X_{\ell m}^\nu=\sum_{jk}\beta_{jk}^\nu\sqrt{\frac{4\pi}{N_j}}h_\ell^jY_{\ell m}^*(n_{jk}) .
\end{align}

The ILC estimate of needlet coefficients of the cleaned map is obtained as a linearly weighted sum of the needlet coefficients
\begin{align}
\beta_{jk}^{\mathrm{NILC}}=\sum_{\nu}\omega_{jk}^{\nu}\beta_{jk}^{X,{\nu}},
\end{align}
here, the requirement to preserve the CMB signal during the cleaning is formulated as a constraint:
\begin{align}
\sum_{\nu}\omega_{jk}^{\nu}=1.
\end{align}
The needlet ILC weights can be calculated by minimizing variance.  The resulting needlet ILC weights that minimise the variance of the reconstructed CMB are expressed as
\begin{align}
\boldsymbol{w}_j^{\mathrm{NILC}}(n_{jk})=\frac{\boldsymbol{\hat{C}}_{jk}^{-1}\boldsymbol{1}}{\boldsymbol{1}^T\boldsymbol{\hat{C}}_{jk}^{-1}\boldsymbol{1}},
\end{align}
with
\begin{align}
 \boldsymbol{\hat{C}}_{jk}=C_{jk}^{\nu_1\times\nu_2} = \langle \beta_j^{\nu_1}(n_{jk})\beta_j^{\nu_2}(n_{jk})\rangle
\end{align}
$\boldsymbol{1}$ is a column vector of all ones.
The NILC cleaned map is transformed from the cleaned needlet maps according eq. (\ref{eq:inver}),
\begin{align}
\hat{X}_{\ell m}^{\mathrm{NILC}}=\sum_{jk}\beta_{jk}^{\mathrm{NILC}}(n_{jk})\sqrt{\frac{4\pi}{N_j}}h_\ell^jY_{\ell m}^*(n_{jk}),
\end{align}

\cite{Zhang:2022} and \cite{Dou:2023} has already demonstrated the effectiveness of NILC in directly removing foregrounds from E and B-mode maps. Here, we adopt the same approach as theirs. Firstly, the CMB polarization maps need to be decomposed into (E, B) maps. When performing spherical harmonic transforms on a partial sky, the orthogonality of spherical harmonics is no longer satisfied, leading to EB leakage. For ground-based CMB observations, this leakage is inevitable, necessitating the correction of EB leakage. Here, we briefly outline the correction for EB leakage, and detailed information can refer to previous work \citep{Liu:2019}. Firstly, decompose the maksed observation polarization maps $(Q, U)$ into $(E, B)$ maps. Then, the inverse transformation of the $(E, 0)$ maps yields $(Q_E, U_E)$ maps. Thirdly, decompose masked $(Q_E, U_E)$  maps to obtain the $B^{\prime}$ maps, which serve as the the leakage template for $B$ map. Finally, we can remove the EB leakage from the masked $B$ map by linear fitting.

The data utilized here are obtained from the simulations in Section \ref{CMB_S4_Experiment}, where we simulated the observational sky maps with the performance of the CMB-S4 experiment. We first decompose the observed polarization (Q, U) maps into (E, B) maps and correct for EB leakage. Subsequently, we apply the NILC method to the multi-frequency observed E and B-mode maps individually to obtain a foreground-cleaned CMB map. It should be noted that there are still residual noise in the foreground-cleaned CMB map. As demonstrated in Section \ref{Noise_Case}, the noise bias in the power spectrum can be  effectively removed through cross-correlation between two HS maps. Furthermore, the noise bias can also be eliminated by estimating the power spectrum of residual noise. Here we adopt the latter method to mitigate the bias of noise. Firstly, 100 noise maps are simulated. Subsequent to applying the NILC method to the multi-frequency sky maps, we keep the NILC weights unchanged and feed these 100 noise sky maps into the NILC, generating 100 residual noise maps. The noise biases in the power spectrum is properly corrected by subtracting the average power spectrum of these 100 noise residual maps from the foreground-cleaned CMB power spectrum.

The Figure \ref{figure_s4_denoise_NILC} displays the power spectra of the foreground-cleaned CMB map. It is evident that the NILC technique accurately recover the EE power spectrum at angular scales $\ell<900$. In contrast, our CNN method can accurately recover the EE power spectrum at finer angular scales. Concerning the lensed B-mode power spectrum, NILC demonstrates the ability to recover the B-mode power spectrum at angular scales $\ell<400$, whereas our approach excels in precisely retrieving the B-mode power spectrum at angular scales $\ell<800$, with reduced errors in the restoration process.

Notably, precise recovery of the B-mode power spectra at angular scales $\ell<300$ is sufficient for detecting primordial gravitational waves. Thus, ILC remains an excellent blind source separation algorithm. Enhanced recovery of smaller-scale  power spectra is beneficial for other cosmological investigations. Lastly, as shown in equation (\ref{eq:conv}), NILC necessitates highly accurate beam modeling, whereas our CNN method focuses on pixel space distribution characteristics within the map, without specific beam requirements.

\section{CONCLUSIONS}
This paper presents the utilization of a machine-learning technique, namely CMBFSCNN \citep{Yan:2023c}, for the extraction of CMB signals from diverse sources of polarized foreground contamination. Our methodology consists of two sequential steps: (1) employing a CNN network to eliminate foreground contamination from the observed CMB map; and (2) employing a cross-correlation technique to mitigate the impact of instrumental noise on the power spectra.

We first implement our pipeline on simulated data designed to the performance of the CMB-S4 experiment. The data simulation incorporates the lensing effect, and our important objective is to accurately recover the weaker lensing BB power spectrum. At the map level, CMBFSCNN effectively eliminates polarized foreground components from both the Q and U maps. The mean absolute deviation (MAD) values between the recovered maps and the corresponding target noisy maps are $0.016 \pm 0.008\ \mu$K for the Q map recovery and $0.021 \pm 0.002\ \mu$K for the U map recovery. Subsequently, we partition the data into two HS maps and perform cross-correlation to mitigate the noise effects on the power spectrum. Notably, the recovered CMB EE power spectra obtained through our methodology closely match the input fiducial CMB information. Additionally, the CMB lensing B-model power spectrum can be accurately recovered at angular scales of $\ell<800$.

Subsequently, we employ this pipeline on full-sky simulated data, emulating the performance of the liteBIRD experiment.  To mitigate the noise level in the network output map, we employ the Internal Linear Combination (ILC) method, which involves obtaining a weighted sum of the noise maps corresponding to the ten frequency bands. The resulting ILC noise plus the beam-convolved CMB map, serves as the training target for the network. At the map level, the CMBFSCNN effectively removes polarized foreground components from both the full sky Q and U maps. The MAD values between the recovered maps and the corresponding target noisy maps are $0.029 \pm 0.004\ \mu$K for the Q map recovery and $0.032 \pm 0.009\ \mu$K for the U map recovery. Following the denoising step, the recovered CMB EE power spectrum closely match the input fiducial CMB information. The CMB lensing B-model power spectrum can be accurately recovered at angular scales up to $\ell<600$. These results suggest that the CMBFSCNN is capable of successfully handling full-sky polarized maps.

Our findings demonstrate the inherent challenge faced by network models in accurately reconstructing pure CMB polarized signals from observed data that is contaminated by diverse foreground sources. However, we remain optimistic about future endeavors, where we endeavor to deepen our understanding and tackle this issue, making substantial progress in this research field. Encouragingly, the network model exhibits the capability to effectively recover the CMB polarized signal plus instrumental noise.

Finally, we illustrate the dependency of our approach on the foreground models. When the actual sky observations align with the simulations in the training data, our outcomes exhibit high quality. Conversely, discrepancies between the real observed signal and the training set simulations result in an increased presence of residual foregrounds in our reconstructed CMB maps. These residual foreground components have an impact on the reconstruction of the lensing B-mode power spectrum. This underscores the necessity of possessing prior knowledge regarding the sky signal and employing this prior information for precise modeling of the sky signals. Further quantitative research on the dependency of our method on the foreground models is left for future work.

We have shown that the CNN method has a good performance in processing CMB polarized maps. More interestingly, the CNN method could be used to reconstruct the foregrounds. We will investigate these interesting issues in future works.

\section{ACKNOWLEDGEMENT}
J.-Q.X. is supported by the National Science Foundation of China, under grant Nos. 12021003, by the National Key R\&D Program of China, Nos. 2020YFC2201603, by the Fundamental Research Funds for the Central Universities. Some of the results in this paper have been derived using the HEALPix (page: \url{https://healpix.sourceforge.io/})

\end{document}